\documentstyle[aps,epsfig,multicol]{revtex}

\newcommand{\br}{{\bf r}}
\newcommand{\bx}{{\bf x}}
\newcommand{\by}{{\bf y}}
\newcommand{\bn}{{\bf n}}
\newcommand{\bp}{{\bf p}}

\newcommand{\bxp}{{\bf x}}
\begin{document}

\draft

\title{Quantum interference and the formation of the proximity effect 
in chaotic normal-metal/superconducting structures}

\author{D. Taras-Semchuk and Alexander Altland}

\address{Theoretische Physik III, Ruhr-Universit\"at-Bochum,
  44780 Bochum, Germany
}

\maketitle                                 
\begin{abstract}
We discuss a number of basic physical mechanisms relevant to the
  formation of the proximity effect in superconductor/normal metal
  (SN) systems. The physics of
  this effect, most notably the phenomenon of density of states
  depletion in normal metals adjacent to superconductors, sensitively
  depends on various aspects of quantum interference and chaotic
  charge carrier dynamics in mesoscopic systems. 
  Specifically, we review why the proximity effect
  sharply discriminates between systems with integrable and chaotic
  dynamics, respectively, and how this feature can be incorporated into
  theories of SN systems. Turning to less well investigated terrain,
  we discuss the impact of quantum diffractive scattering on the
  structure of the density of states in the normal region.  We
  consider ballistic systems weakly disordered by pointlike impurities
  as a test case and demonstrate that diffractive processes akin to
  normal metal weak localization lead to the formation of a hard
  spectral gap -- a hallmark of SN systems with chaotic dynamics.
  Turning to the more difficult case of clean systems with chaotic
  boundary scattering, we argue that semiclassical approaches,
  based on classifications in terms of classical trajectories, cannot
explain the gap phenomenon. Employing an alternative formalism based
  on elements of quasiclassics and the ballistic $\sigma$-model, we
  demonstrate that the inverse of the so-called Ehrenfest time is
  the relevant energy scale in this context. We discuss some
  fundamental difficulties related to the formulation of low energy theories
  of mesoscopic chaotic systems in general and how they prevent us
  from analysing the gap structure in a rigorous manner. Given these
  difficulties, we argue that the proximity effect
  represents a basic and challenging test phenomenon for theories of
  quantum chaotic systems.
\end{abstract}


\begin{multicols}{2}
\narrowtext

\section{Introduction}
\label{intro}

Superconductors attached to normal metals tend to export aspects of
their anomalous properties into the adjacent normal metal
region. Termed the proximity effect, this tendency
manifests itself in phenomena such as the
Josephson effects, anomalous charge and
thermal transport coefficients, and altered
thermodynamic properties (see
e.g. Refs.\cite{Bee:RMP,Zaikin88,Lambert98} for extensive reviews).  
All these effects find their
common origin in the fact that a superconductor pairing field
amplitude may penetrate deeply into the normal metal region, before decaying  
due to the absence
of attractive interaction mechanisms.  

Perhaps the most direct manifestation of the presence of a finite
pairing field amplitude in a normal metal adjacent to a superconductor
is a massive suppression of the single particle density of states
(DoS). This phenomenon, first noticed almost forty years ago, has a
long history of research.  Beginning with a classic paper of de
Gennes and Saint-James\cite{gennes63}, the normal metal DoS of a wide
family of superconductor--normal-metal hybrid systems (SN-systems) has
been investigated. Among these are superconductors coated with normal
metal thin films, both clean and disordered\cite{gennes63,mcmillan68},
bulk diffusive compounds in contact with superconductor terminals
\cite{golubov88,golubov89,Zhou95,belzig96,altland00}
and various types of SN quantum dot structures\cite{Bee:RMP,altland96,melsen96,melsen97,oreg99,volkov95,clerk00}.

In
general the {\it quantitative} functional form of the energy dependent
DoS, $\nu(\epsilon)$, of any of these systems depends on the system
category, and on a number of relevant material parameters.
Interestingly, however, a simple ordering principle exists, as first
pointed out by Melsen {\em et al}.\cite{melsen96,melsen97}, according
to which the {\it qualitative} structure of the DoS falls into one of
only two different categories: For systems whose normal compound is
{\it integrable} -- in the sense that the dynamics of the single
particle degrees of freedom becomes integrable in the classical limit
-- the DoS vanishes in a close to linear 
fashion\cite{gennes63,melsen96,melsen97} as the excitation
energy $\epsilon$ approaches zero. Of course, due to
the inevitable presence of imperfections, truly integrable systems in
condensed matter physics do not exist. However, a number of systems,
e.g. thin clean metallic films, come reasonably close to the
integrable limit implying that the proximity DoS obeys a continuously
decreasing behavior down to very small excitation energies
$\epsilon$. In contrast, the DoS of genuinely {\it chaotic} systems -- 
significantly disordered systems or chaotic quantum dot 
structures -- is gapped: $\nu(\epsilon)$ falls to zero at some
{\it finite} threshold energy $\epsilon^\ast$. 
For energy values $|\epsilon|$ smaller than $\epsilon^\ast$, the DoS
is strictly vanishing. 

Beyond its practical relevance in the physics of SN systems, this
twofold categorization scheme is of noticeable conceptual interest.
Indeed, the observation that a quantity as elementary as the single
particle DoS of may serve as an indicator sharply discriminating
between integrable and chaotic dynamics has led to a wave of interest
in the interplay between superconductor correlations and quantum
chaotic
dynamics\cite{altland96,melsen96,melsen97,lodder,Schomerus99,Kosztin95,Brouwer96,Richter}.  
Unfortunately,
the established tools of theoretical mesoscopic superconductivity --
quasiclassical Green function methods and diagrammatic techniques --
are not straightforwardly applicable to clean chaotic SN structures.
As an alternative to direct numerical analyses, an approach based on
classical path counting, augmented by a Bohr-Sommerfeld quantization
rule, has become popular.  Referring to a more substantial discussion
to below, we here merely note that the Bohr-Sommerfeld (BS) approach
readily produces a massive suppression of the low energy DoS of
chaotic systems\cite{lodder,Schomerus99} 
(as compared to the integrable case.) Furthermore, the
energetic extension of the depletion region turns out to be in good
accord with the results obtained by direct
diagonalization\cite{melsen96,melsen97,lodder,Schomerus99,Richter}. 
In this sense, the BS
quantization scheme represents a comparatively simple and widely
applicable tool for exploring the DoS of SN systems.

There is, however, one key feature of the DoS gap that is missed by
the BS-scheme: for chaotic systems the gap tends to be {\it hard} (in
the sense that $\nu(\epsilon)$ strictly vanishes for energies
$|\epsilon| < \epsilon^\ast$.) Hard gaps are found for a wide class of
chaotic SN systems: For the particular case of diffusive SN systems,
the structure of a hard gap has been obtained analytically by
quasiclassical
methods\cite{usadel70,golubov88,golubov89,Zhou95,belzig96,altland00}.
A numerical study by Pilgram {\em et al.}\cite{pilgram} has shown that
the gap of weakly disordered systems, i.e. systems whose spatial
extent is smaller than the elastic mean free path, is also hard
(within, of course, the numerical accuracy). The same applies to the
numerically obtained DoS profiles of systems with smooth chaotic
boundary scattering displayed in Ref.\cite{lodder}. Finally, chaotic
systems that are completely ergodic, in the sense that their quantum
mechanics can be modelled by random matrix theory, also exhibit a hard
gap\cite{melsen96,melsen97}.

In contrast, the very construction of the BS scheme implies that it
categorically produces a {\em finite} (albeit small) DoS at non-zero
energies\cite{melsen96,melsen97,lodder,Schomerus99,Richter}. 
That no gap is seen has to do with the fact that the the BS
approach does not fully account for the quantum mechanical nature of the
quasi-particle propagation in an SN environment.  More specifically,
the elimination of {\it all} states below a certain threshold
energy is caused by mechanisms of quantum diffraction similar to those
responsible for weak localization corrections in normal metals.  The
essential {\it difference} to the weak localization processes
operative in normal mesoscopic systems is that diffraction processes
in SN systems are not small. Indeed it is one of the main objectives
of the present paper to show that at low enough energies these
`corrections' diverge and that a non-perturbative treatment is needed
to correctly obtain the gapped DoS.

Qualitatively, the results of our analysis may be summarized as
follows: in chaotic SN systems, there are at least two different time
scales that are of relevance for the low energy DoS. The first is
the average classical time $t_{\rm c}$ it takes to establish contact
with the superconductor. This scale can be set by be the diffusion
time time through an extended disordered system, the inverse tunneling
rate through a nearly insulating superconductor normal metal
interface, or the average ballistic time of flight through a clean
quantum dot. The second time scale is, in general, of quantum
mechanical nature, and is the time $t_{\rm d}$ a minimal wave package
takes to lose memory of its classical initial conditions. In a medium
with hard ($s$-wave) scatterers, $t_{\rm d}$ will coincide with the
classical scattering time $\tau$. In systems with smooth scattering
potentials -- no matter whether disordered or clean -- $t_{\rm d}$ is
determined by the Ehrenfest time $t_{\rm E}$\cite{AlLa1}, that is to be
discussed extensively below.  For the structure of the low energy DoS,
the larger of these two timescales is of foremost relevance. For
example, in a typical diffusive system with hard impurities, the
scattering time is much shorter than both the diffusion time and the
tunneling time into the superconductor. In this case, $t_{\rm c} >
t_{\rm d}$ and the parametric scale determining the gap
edge\cite{golubov88,Zhou95,belzig96,altland00} is $t_{\rm c}^{-1}$.

In this paper the focus will be on the complementary class of systems
with $t_{\rm d} \gg t_{\rm c}$. We will begin by considering the
prototypical class of ballistic systems weakly polluted by some
$s$-wave impurities and well coupled to a superconductor.  For such
systems the scattering time is larger than the ballistic time of
flight (and the tunneling time to the superconductor) and the above
condition is met.  Operating within a synthesis of the quasiclassical
formalism and the ballistic non-linear
$\sigma$-model\cite{KhM2,A3S:Nph}, we will identify the relevant
quantum diffraction processes, and show how these contributions lead
to a vanishing of the DoS for energies $\epsilon < \epsilon^\ast
\sim \hbar/(2\tau)$.  

Building on the insight gained in this analysis, we will then proceed
to the more complex problem of clean chaotic systems with smooth
potentials. In this case, the relevant corrections to the classical
picture emerge from an interplay of quantum wave packet spreading and
the exponential trajectory divergence inherent to chaotic dynamics.
The characteristic time scale for these processes is the
aforementioned Ehrenfest time. Below we will demonstrate that the
Ehrenfest time is the relevant energy scale for the DoS of chaotic
systems with smooth scattering.  It should be mentioned right away
that the quantitative parts of this part of the analysis will include
certain assumptions and, therefore, do not attain the status
of a rigorous calculation. Further, our formalism is not capable of
exploring structures in the close vicinity of the estimated
gap edge $\sim \hbar/t_{\rm E}$. Notwithstanding these
deficiencies, our analysis demonstrates the suppression of the DoS
below its semiclassically estimated value. Although we cannot
mathematically prove that this quantum suppression mechanism produces
a hard gap, the close structural analogy with the hard scattering
problem makes us believe that this is the case. That a hard gap of the
order of the inverse Ehrenfest time should be generated has already
been conjectured by Lodder and Nazarov\cite{lodder}.  

We would like to point out that the phrase 'hard gap', as used in this
paper, needs some qualification: refined analysis of the spectral
structure of diffusive Andreev systems has shown the existence of a
tail below the quasiclassical gap
edge\cite{skvortsov} that is exponentially 
small in $g$, where $g \gg 1$ is the dimensionless conductance
of the normal
component of the system.
Similar contributions have been indentified in billiards with
randomly fluctuating geometry\cite{vavilov}. In diffusive systems, the
formation of these anomalous `tail' states is due to exponentially
rare potential configurations which lead to an effective decoupling of
the superconductor and normal metal\cite{altland00}. An interesting
question, not addressed in the present paper, is whether similar
states exist in clean chaotic environments. It should be emphasized,
though, that tail state formation is a mechanism conceptually
independent from the quantum diffractive suppression of the
semiclassical results discussed in this paper.

Finally, let us mention that much of the conceptual ground on which
our analysis of the DoS in chaotic systems with smooth boundaries is
based was laid in a seminal paper by Aleiner and Larkin\cite{AlLa1}
(AL). In this work the fundamental importance of the Ehrenfest time
for the formation of weak localization corrections in normal chaotic
systems was realized and substantiated.  Many of the ideas first
formulated by AL have been included in our analysis of the
superconductor problem.

The paper is organized as follows. In section \ref{sec:qualitative},
we discuss some general physics associated with Andreev reflection in 
SN structures, introduce the BS approach and discuss the role of
quantum diffraction processes in establishing the gap in the spectrum
of the normal region. In section \ref{sec:qc} we discuss how the
BS approach may be recovered from the quasiclassical Eilenberger
equation in the diffractionless limit, and establish a simple trial
solution of the quasiclassical equations in the presence of a 
collision operator. In section \ref{sec:sigma} we introduce a
functional integral formulation for the disordered Green function of
an SN system in terms of a ballistic sigma-model.
In section \ref{sec:ballistic_box} we show how this
formalism may be applied to determine the spectrum of a SN system with 
bulk, ballistic $s$-wave disorder, and establish the presence of a hard gap
of the order of the inverse scattering time. 
In section \ref{sec:clean_chaotic} we 
turn to the study of the DoS in clean chaotic SN systems, in the
presence of a smooth
scattering potential, and discuss the appearance of the inverse
Ehrenfest time in the structure of the DoS. 
Section \ref{sec:discuss} concludes with a discussion and summary.

\section{Qualitative Physics of the Gap}
\label{sec:qualitative}

The basic microscopic mechanism of coupling between a 
 superconductor and an 
adjacent normal metal is Andreev reflection\cite{andreev65}:
a normal metal electron impingent on the superconductor interface is
converted into a hole (charge balance being restored by the
formation of a superconductor Cooper pair) which is reflected back
into the normal metal. This type of scattering has a number of
peculiar properties which are at the root of the formation of 
the DoS gap and which are summarized below for 
later reference: 
\begin{itemize}
\item Andreev reflection is a retro-reflection, i.e. the 
  hole is essentially reflected back into the path of the incident
  electron where
\item there is a slight angular mismatch set by the excitation energy
  $\epsilon$ of the electron above the Fermi energy $E_F$.
\item For an electron of energy $\epsilon$, the reflected hole has
  energy $-\epsilon$.
\item the hole acquires a scattering phase, which takes the value of  
  $\pi/2-\phi$ (for an excitation energy $\epsilon$ far below the
  superconducting gap), where $\phi$ is the value of the superconducting
  order parameter at the point of incidence.
\end{itemize}

\begin{figure}[h]
  \centerline{\epsfxsize=2.5in
  \epsfbox{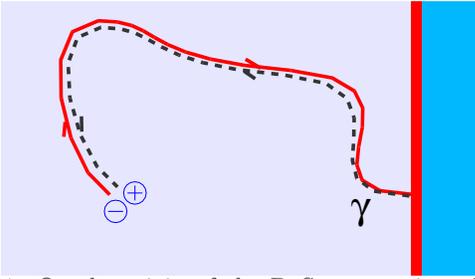}
}
  \caption{On the origin of the DoS suppression. The
    coupling to a superconductor (dark shaded region) leads to the formation of a stable
    pair amplitude $\sim \langle\psi\psi\rangle$ in the normal metal
    (light shaded region).}
  \label{fig:proximity_mechanism}
\end{figure}

From these properties, the origin of the DoS suppression in normal metals
adjacent to superconductors may be readily understood. Assume that an
electron of some small excitation energy $\epsilon$ is injected
somewhere into the normal metal (at the beginning of the solid line in
Fig.~\ref{fig:proximity_mechanism} say.) The electron will travel
along some path $\gamma$ in the normal metal and at some point hit the
superconductor where it is converted into a hole. The hole then
propagates approximately (up to the above mentioned angular mismatch)
back along the path $\gamma$ (dotted line in 
Fig.~\ref{fig:proximity_mechanism}) and eventually arrives in the
vicinity of the point of injection. Seen as a whole, this process
amounts to the formation of a local net pair amplitude $\sim \langle
\psi^\dagger(\epsilon) \psi^\dagger(\epsilon) \rangle$, i.e. an
amplitude for the creation of an electron and an annihilation of a
hole, in the normal metal. Importantly, this induced Cooper pair
amplitude is not weighted by rapidly fluctuating quantum mechanical
phases. To understand this point, notice that the quantum phase 
of the  incident electron is 
essentially given by the classical action $S_\gamma(\epsilon)$ for
traversal along $\gamma$. The
action of the Andreev reflected hole (a missing electron) is given by
$-S_\gamma(-\epsilon)$. These two contributions largely cancel against 
each other, implying that a slowly fluctuating net amplitude obtains. 

Unitarity implies that the formation of a non-vanishing Cooper pair
amplitude $\sim \langle \psi(\epsilon) \psi(\epsilon) \rangle$ must be
compensated for by a diminished 'normal state' amplitude $\langle
\psi^\dagger(\epsilon) \psi(\epsilon) \rangle$ or, in other words, by
a diminished local single particle density of states (see below for a
more precise phrasing of this argument.) It is also clear that this
suppression mechanism can only be effective for small energies
$\epsilon$, and for spatial regions not too distant from the
superconductor: For increasing $\epsilon$ or, equivalently, increasing
length of the characteristic paths $\gamma$, the action difference
$S_\gamma(\epsilon) - S_\gamma(-\epsilon)$ increases, electron and
hole run out of phase and the induced pair amplitude randomizes to
zero.

\begin{figure}[h]
  \centerline{\epsfxsize=2.5in
  \epsfbox{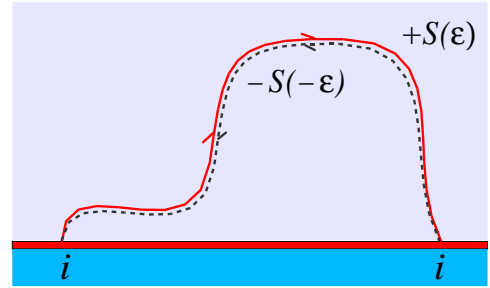}
}\vspace{0.5cm}
  \caption{Generic structure of the classical trajectories forming the 
    basis of the BS scheme.}
 \label{fig:bohr_sommerfeld_idea_II}
\end{figure}

This mechanism of `action cancellation' is the first of two basic
physical principles determining the phenomenology of the proximity
effect. It not only explains the formation of a pair amplitude but
also the well known sensitivity of the DoS suppression to the
classical integrability properties of the system. This
latter point is most easily understood within the Bohr-Sommerfeld
approach to the problem. The BS approach is based on the condition
that the classical action of any closed trajectory traced out by the
constituent particles of the system be an integer multiple of $h$. The
generic topology of closed trajectories of an SN-system is displayed
in Fig.~\ref{fig:bohr_sommerfeld_idea_II}: for sufficiently small energies,
$\epsilon$, an electron/hole trajectory encompassing two Andreev
events closes into itself. The total action of such a trajectory is
given by
$$
S_\gamma(\epsilon) - S_\gamma(-\epsilon) -{h\over 2},
$$
where the term $-h$ is due to the Andreev scattering phase shift
$i\exp(i\phi)$\cite{fn1}. Expanding $S_\gamma(\epsilon) = S_\gamma(0)
+ T_\gamma \epsilon$, where $T_\gamma$ is the time of traversal of
$\gamma$, one obtains the condition
$$
\frac{2 T_\gamma \epsilon}{h} \stackrel{!}{=}  n+{1\over 2},
$$
where $n$ is an integer. This condition implies that for small
$\epsilon$ the structure of the DoS is determined by the distribution
$P(L_\gamma)$ of classical paths of large length $L_\gamma = v_F
T_\gamma$. (See Refs.\cite{melsen96,melsen97,lodder,Schomerus99} 
for the precise
formulation of the BS evaluated DoS.) For classically integrable and
classically chaotic systems, respectively, these distributions are
qualitatively different. 
More specifically, for open chaotic systems,
$P(L_\gamma)\propto \exp(-L_\gamma/(v_F t_{\rm f}))$ is exponentially
small (where $t_{\rm f}$ is the escape time, see section \ref{sec:qc})
whereas for open integrable systems $P(L_\gamma)$ decays only
algebraically. (Here the attribute `open' means that we are dealing
with systems that are connected to outside leads.)  The exponential
rareness of long trajectories provides the BS
explanation\cite{Schomerus99} for the
smallness of the low energy DoS in chaotic systems.

Notice, however, that irrespective of the structure of the system,
there will {\it always} be some finite probability $P(L)$ of finding
trajectories of length $\sim \hbar v_F/\epsilon$.  In other words,
although the BS prediction for the DoS may be small, it will {\em never} be
strictly vanishing (for any finite $\epsilon$.)  For chaotic systems,
this is in contradiction with the results obtained by other methods.
Specifically, for diffusive metallic
systems\cite{golubov88,golubov89,Zhou95,belzig96,altland00,golubov95,altland98},
weakly disordered ballistic systems\cite{pilgram}, and
'zero-dimensional' quantum dot type
systems\cite{melsen96,melsen97} a true DoS gap (strictly
vanishing DoS for small energies) has been obtained by
quasi-classical, numerical, and random-matrix  methods, respectively.

\begin{figure}[h]
  \centerline{\epsfxsize=2.0in
  \epsfbox{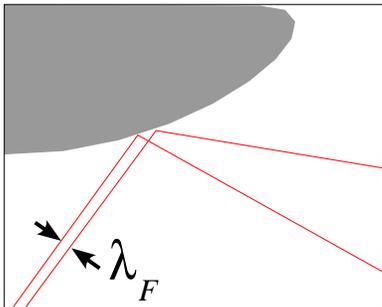}}
  \caption{Quantum diffraction in a chaotic system: after a time set
    by the largest Liapunov exponent of the system, two initially
    close trajectories diverge.}
  \label{fig:diffraction}
\end{figure}

What is the reason for the discrepancy between these results and the BS
approach? The answer is that the latter does not account for the
second key mechanism relevant for the physics of the proximity effect,
{\it quantum diffraction}: the discussion above was based on a
representation of transition amplitudes in terms of idealized
classical trajectories. However, in a quantum system the uncertainty
principle imposes an upper cutoff $\hbar$ on the phase space
resolution. In practical terms this means that the propagation along
the trajectories entering the BS counting scheme must be interpreted
as a caricature of the propagation of quantum minimal wave packets.
The transverse spreading of these packets is at least of the order of
the Fermi wave length $\lambda_F$ (cf. Fig.~\ref{fig:diffraction}).

Whereas in integrable systems this quantum spreading is of no real
concern, its consequences in chaotic systems are profound. The
existence of a globally positive Liapunov exponent $\lambda$ entails
that a microscopic initial uncertainty of $O(\lambda_F)$ increases
exponentially $\sim \lambda_F \exp(t \lambda)$ and, after a time
$t_{\rm E} = \lambda^{-1} \ln(a/\lambda_F)$, reaches the scale $a$ of
the macroscopic geometry of the system. On time scales beyond this
so-called Ehrenfest time, a classification scheme based on classical
trajectories becomes meaningless.  The consequences of this spreading
of minimal wave packages have been explored in AL. In that paper a
mechanism was discussed by which the spreading leads to the formation
of weak localization type quantum corrections to the classical (Drude)
conductance of chaotic systems: in the terminology of semiclassics,
two initially close trajectories split up but eventually recombine to
give an overall phase-insensitive weak localization correction.

A similar mechanism affects the mean DoS of SN systems. However,
unlike the weak localization corrections to the normal conductance --
generally a small correction to a large classical background --
quantum diffraction in chaotic SN systems is of vital relevance for
even the 
leading order, {\it qualitative} structure of the DoS. In
particular, it is responsible for the strict vanishing of
$\nu(\epsilon)$ below a certain threshold energy.

\begin{figure}[h]
  \centerline{\epsfxsize=2.0in
  \epsfbox{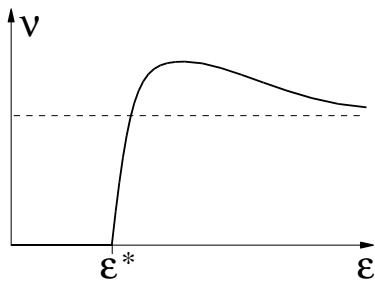}}
  \caption{Qualitative structure of the mean DoS of a compact diffusive system 
    in contact with a superconductor.}
  \label{fig:mean_dos}
\end{figure}

For a particular system class, such as diffuse disordered systems, the
effect of these quantum contributions can be investigated in
quantitative detail.  The idea is to compare the trajectory oriented
picture outlined above with quantitative solutions obtained by other
methods. Indeed, the mean DoS of of a large variety of SN systems with
diffusive N compounds has been obtained by solving the quasiclassical
Usadel
equations\cite{usadel70,golubov88,golubov89,Zhou95,belzig96,altland00,fn2}.
Mathematically, the Usadel equations  are second order nonlinear
differential equations for the average Green function of the system. They
are obtained by a quasiclassical expansion of the microscopic Gorkov
equations, with self consistent account for the (quantum) scattering
off impurities. Solving these equations it has been found that for
systems with an N region of finite extension $L$, the DoS profile
universally contains a hard gap of the type schematically drawn in
Fig.~\ref{fig:mean_dos}. We next address the question of how this
phenomenon can be reconciled with the semiclassical picture discussed
above.

For concreteness, let us consider a diffusive system in the quantum
dot regime, i.e. a regime where the Thouless energy $E_c = \hbar
D/L^2$ (where $D$ is the diffusion constant) is the largest energy
scale in the problem.  (The discussion can be adapted
straightforwardly to other types of system configurations.) In this
case, the Usadel equation is controlled by the dimensionless parameter
$\kappa \equiv \gamma/\epsilon$, where $\gamma$ is an energy scale
representing the coupling strength to the superconductor.  Although
the parameter $\kappa$ need not be small, it is nevertheless
instructive to expand the Usadel equation in $\kappa$; not to obtain
quantitative results, but rather to understand the basic physical
mechanism responsible for the buildup of the gapped DoS.  Referring to
Ref.\cite{altland00} for a more detailed discussion we here merely
sketch the qualitative picture.

\begin{figure}[h]
  \centerline{\epsfxsize=3.5in \epsfbox{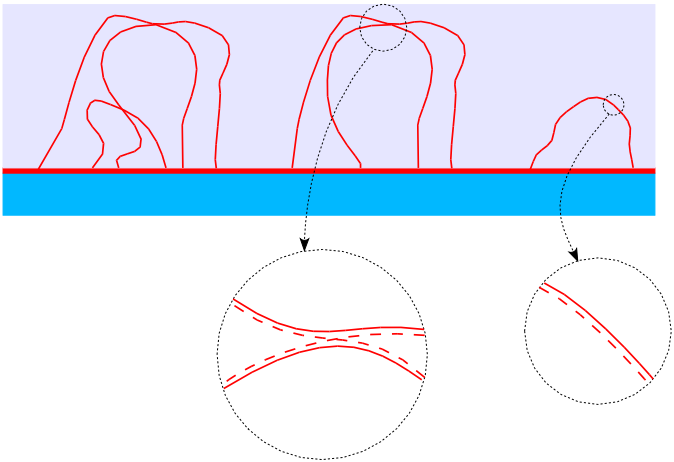}}
  \caption{On the semiclassical interpretation of the solution of the
    Usadel equations.}
  \label{fig:hikami}
\end{figure}

The linearity of $\kappa$ in both the coupling to the superconductor
and in $1/\epsilon$ means that each power of this parameter
corresponds to a single Andreev scattering event, coupled to a
diffusion mode\cite{fn3}.  The diffusion mode describes the coherent
propagation of a particle and a hole through a diffusive background
(Fig.~\ref{fig:hikami}, right inset), i.e. it is the diffusive
implementation of the interfering particle/hole amplitudes discussed
above.  A cartoon of the correction to the DoS of the lowest order in
$\kappa$ is depicted in Fig.~\ref{fig:hikami}, right.  Technically,
this correction is described by a single diffusive ladder diagram with
two Andreev scattering vertices\cite{altland00}. However, as our
present discussion serves a purely qualitative purpose, we are not
going to elaborate on the diagrammatic interpretation any further. The
only thing to be kept in mind is that a perturbative
semiclassical/diagrammatic approach to calculating the DoS amounts to
an expansion in powers of the parameter $\kappa$.

Already at the next to leading order in $\kappa$
contributions of a different topology appear (Fig.~\ref{fig:hikami},
center). The distinguishing feature of these corrections is that they
contain `junction' points at which a pair of interfering particle/hole
amplitudes splits up (but later recombines to form an overall phase
coherent correction). These corrections
\begin{itemize}
\item are {\em not} included in a semiclassical approach based on a
  classification in terms of purely classical trajectories, and
\item are structurally similar to the weak localization corrections of 
  normal metals.  In fact, the basic junction region (left inset in
  Fig.~\ref{fig:hikami}) is a real space representation of what in the
  context of disordered systems is called a Hikami box. Within
  diagrammatic approaches these objects appear as vertices in the
  diagrams describing metallic weak 
  localization. 
\item Finally,
  the diffractive splitting of a coherent particle/hole amplitude is a 
  process  that does not hinge on the diffusivity of
  the medium, but merely on the chaoticity of the underlying
  dynamics. This point will be discussed in great detail in
  section \ref{sec:clean_chaotic} below. 
\end{itemize}

The main difference to standard  weak localization is that 
the DoS `corrections'
depicted in Fig.~\ref{fig:hikami} are not small.  In fact each
additional junction introduces two more powers of $\kappa$, and this
parameter can assume arbitrary values. As $\epsilon$ is lowered and
$\kappa \nearrow 1$,  a situation arises where diffraction processes of
infinite order have to be taken into account, a task which can clearly
cannot be accomplished within a semiclassical trajectory oriented
approach. In fact it turns out that
for $\kappa \ge 1$  
expansions in terms of trajectories (diagrams) do not even
converge in principle. 

To understand the origin of the problem, it is instructive to compare
the semiclassical expansion with the explicit solution of the Usadel
equation. For the quantum dot type system presently under
consideration, the quasiclassical expression of the 
DoS\cite{melsen96,melsen97,oreg99,volkov95,clerk00} is given by
the simple expression,
$$
\nu(\kappa) = \nu_0 \, {\rm Re\,}{1\over \sqrt{1-\kappa^2}},
$$
where $\nu_0$ is the bare metallic density of states (in the absence 
of a superconductor). 
For our purposes, the most important feature of this expression is
its BCS type essential singularity at $\kappa=1$. The structure of
this singularity implies that (a) $\kappa=1$ marks the DoS gap, for
(b) values $\kappa>1$ the DoS strictly vanishes and (c) the gap
region $\kappa>1$ is out of reach of a semiclassical/diagrammatic
perturbative expansion centered around the metallic limit $\kappa\to
0$. For other types of diffusive systems, the relevant control
parameter may be a different one. Also the analytical structure of the
quasiclassical DoS will in general be way more complicated than the
expression above\cite{golubov88,golubov89,Zhou95,belzig96,altland00}. However
the three points (a)-(c) remain generally valid.

In the rest of the paper we will try to explore to what extent these
structures generalize from diffusive to more general system classes.
This analysis will be formulated within a 
hybrid formalism comprising  quasiclassical Green function
methods and elements of the supersymmetry
approach to disordered systems. The relevant theoretical background material is reviewed in
the next two sections.

\section{Quasiclassical Formulation}\label{sec:qc}

To promote the discussion of the previous section to a more
quantitative level, we start out from the Eilenberger
equation\cite{eilenberger68}. While the Eilenberger equation may be
written in a general form that describes either point-like or
long-ranged scatterers, we focus in this section on its application to
systems with hard boundaries and weakly disordered by point-like
scatterers (`disordered billiards'). The case of clean systems with
soft potentials will be discussed more fully in section
\ref{sec:clean_chaotic}.

The general form of the Eilenberger equation may be written as 
\begin{eqnarray}
  \label{eilenberger}
&&  \left(\hbar{\cal  L}  - i[ \sigma_3(\epsilon + \hat \Delta(\br)),\;
]\right)g(\br,\bn) = {\cal O}_{\rm coll}[g],\nonumber\\
&&\qquad g(\br,\bn)g(\br,\bn)=\openone,
\end{eqnarray}
where $g(\br,\bn)$ is the quasiclassical retarded Green function of
the system\cite{eilenberger68,fn4}, ${\cal L}$ the classical Liouville
operator, ${\cal L} = \{ H, \;\;\}$ acting in the phase space of
points $(\br,\bp)$, and $H$ the Hamiltonian function of the system. The
coordinates $\bxp \equiv (\br,\bn)$ parameterize a $2d-1$ dimensional
shell of constant energy $E_F=p_F^2/(2m)$, where $\bn=\bp/p_F$, $p_F$
is the Fermi momentum and $d$ the dimensionality of the system.
Further, $\sigma_i$ are Pauli matrices operating in the two component
particle-hole ({\sc ph})-space, and $\hat \Delta \equiv \Delta(\cos
\phi \sigma_1 + \sin\phi \sigma_2)$ is the Nambu representation of the
phase dependent order parameter. For an SN system the support of $\hat
\Delta$ is limited to the S region where it must, in principle, be
determined self consistently. With regard to the the structure of the
proximity induced gap in the N region, however, the details of the
order parameter profile in S are of secondary importance, and it
suffices to {\it model} $\Delta$ by a step function
$$
\Delta(\br)=\left\{
  \begin{array}{ll}
    \Delta&\br \in {\rm S},\cr
     0    &\br \in {\rm N}
  \end{array}
\right..
$$
For simplicity we will assume that the superconductor has a constant
order parameter phase $\phi=0$.

The operator ${\cal O}_{\rm coll}[g]$ appearing on the right-hand side of
equation (\ref{eilenberger}) describes the effect of impurity
scattering. Its functional form depends on the microscopic realization
of the scattering potential:
\begin{eqnarray*}
{\cal O}_{\rm coll}[g]&=& -{\hbar\over 2}\int \frac{d\bn'}{S_d} 
[w(\bn-\bn')
g(\br,\bn'), g(\br,\bn)], \\
w(\bn-\bn') &=& 2\pi\nu n_i|\tilde{V}(\bn-\bn')|^2,  
\end{eqnarray*}
where 
$S_d$ is the surface of the $d$-dimensional unit sphere. Here
$\tilde{V}$ is the Fourier transform of the scattering amplitude and
$n_i$ is the impurity concentration, while as usual the limits of
$\tilde{V}\to 0$ and $n_i\to\infty$ are implied, such that $n_i |V|^2$ is
constant.  The scattering lifetime $\tau$ is defined
through $w(\bn-\bn')$ by
\begin{eqnarray*}
\frac{1}{\tau} = \int \frac{d\bn'}{2\pi} w(\bn-\bn').
\end{eqnarray*}
For the weakly disordered systems considered in this paper, the
scattering time exceeds the typical time of flight through the system.

For finite-size impurities characterized by some Gaussian potential
envelope function, we have
$$
w(\bn-\bn') \propto \exp\left(-{p_F^2 a^2\over \hbar^2} |\bn-\bn'|^2\right),
$$
where $a$ is the typical extent of the scattering region. In the
remainder of this section we will focus on an ultra-quantum limit
obtained by sending $\hbar$ to infinity {\it at fixed Fermi energy $E_F$}
and fixed $a$. (Notice the differences with the limit of $\hbar
\to 0$ at fixed electron density $\propto (p_F/\hbar)^d$, usually taken
in Boltzmann transport theories.) In this limit the form factor $w$
becomes independent of the momentum transfer $\propto \bn-\bn'$:
hence this limit corresponds to purely $s$-wave scattering. 
The motivation for considering the $s$-wave limit first
is that the impurity scattering becomes maximally diffractive,
making it an ideal test case for studying the
impact of quantum effects on the DoS-profile. Indeed, for $s$-wave
impurities, it is the ordinary scattering time $\tau$ that sets the time scale
after which classical trajectories are destabilized through
diffraction and one expects $\hbar/ \tau$ to be relevant energy
scale for the DoS gap. More complex mechanisms of diffraction 
will be considered in section \ref{sec:clean_chaotic} when we
discuss the proximity effect in chaotic systems.

As a preparation to the discussion of the full Eilenberger equation
(\ref{eilenberger}) it is instructive to consider the diffractionless 
limit, for which $\tau
\to \infty$. Physically, this limit
describes the regime of large energies, $\epsilon \gg \hbar/\tau$ (time
scales much shorter than the scattering time) for which
the collision operator
is negligibly weak as compared to both the order parameter and the
energy argument. In the limit of no diffraction, the Eilenberger
equation reduces to an ordinary first order differential equation that
can be solved analytically\cite{lodder}.  Indeed, each phase space
point $\bxp$ can unambiguously be assigned to a classical trajectory
$\bxp(t)$ passing through it. The initial and final point of this
trajectory are set by two points of Andreev reflection (cf.
Fig.~\ref{fig:bohr_sommerfeld_idea_II}).  Denoting the time
of the first (second) Andreev scattering event by $-(+)t_0/2$ and using that
$\dot \bxp = \{H,\bxp\}$, the Eilenberger equation in the normal
region reduces to
$$
(\hbar d_t - i\epsilon [\sigma_3,\;])g(\bxp(t))=0,\qquad t\in [-t_0/2,t_0/2].
$$
In the limit $\Delta/\epsilon \to \infty$ the solution of this
equation is given by\cite{lodder},
\begin{equation}
  \label{solution_clean}
g(\bxp(t))= \sin \theta \cos\left(  \frac{\epsilon t}{\hbar}\right) \sigma_2
+ \cos \theta \sigma_3 -\sin \theta \sin\left(\frac{\epsilon t}{\hbar}\right)
\sigma_2,  
\end{equation}
where $\cos \theta =- i \tan(\epsilon t_0/\hbar)$. 
This expression contains two pieces of
information that will be of key importance for all what follows: (i)
The non-diffractive Eilenberger formalism is equivalent to the
BS-approach\cite{melsen96,melsen97,lodder,Schomerus99}. 
To derive this equivalence one starts out
from the general formula for the DoS,
$$
\nu(\epsilon) =  {\nu_0\over 2}{\rm Re}\int d\bxp\, {\rm
  tr\,}(g(\bxp,\epsilon^+)\sigma_3),
$$
where $\nu_0$ is the metallic reference DoS and the phase space
integral $\int d\bxp=1$ is normalized to unity. Substituting the clean
solution for $\epsilon^+ = \epsilon +i0$ it is a straightforward
matter to show that
$$
\nu(\epsilon) \stackrel{\tau \to \infty}{=} \pi\nu_0 \int d\bxp
\sum_{n=-\infty}^\infty 
\delta\left({\epsilon t_0(\bxp) \over \hbar} - 
  \left(n+{1\over 2}\right)\pi\right) 
$$
which is the BS result\cite{melsen96,melsen97,lodder,Schomerus99}. (ii) 
For energies $\epsilon \ll \hbar/ t_0(\bxp)$
much smaller than the inverse flight time through the normal segment of the
trajectory, $g(\bxp) \approx \sigma_2$ is approximately 'locked' to the
order parameter of the adjacent superconductor. This fact will become
of importance below when we analyze the full Eilenberger equation. 

We now proceed to the discussion of diffraction effects. For finite
$\tau$, the Eilenberger equations along different trajectories become
coupled and no rigorous solution is possible. In chaotic systems,
however, one may invoke statistical arguments to turn the complexity
of the problem to an advantage. (Notice that under the conditions
specified above, the presence of impurities alone suffices to render
the dynamics chaotic on time scales $t > \tau$.)  More specifically,
we are going to exploit the features of the path length distribution
function $P(v_F t_0)$ of systems with chaotic dynamics to propose a
trial solution of the Eilenberger equation that should be close to the
unknown exact solution. Contrary to the BS result, the DoS derived
from the approximate solution will exhibit a hard gap.  In the next
section we will then show that this gap is robust to all orders of
perturbation theory around the trial solution.

We begin by noticing that in the diffractive problem two intrinsic
time scales exist: The scattering time $\tau$ and the `typical escape
time' $t_{\rm f}$, i.e. the time scale $t_{\rm f}= L_{\rm f}/v_F$
corresponding to the maximum $L_{\rm f}$ of the path length
distribution function $P(L)$. For a ballistic system, we have 
$\tau \gg t_{\rm
  f}$, and so we conclude that three qualitatively different temporal
regimes exist: for short times $t\ll t_{\rm f}$, both diffraction
effects and the differences between integrable and chaotic dynamics
are negligible, and the DoS is well approximated by the BS formula.
For intermediate times $t_{\rm f} \ll t \ll \tau$ diffraction is still
inessential but the structure of the function $P(L)$ begin to matter.
Specifically, the exponential rarity of trajectories with $t_0 >
t_{\rm f}$ implies a strongly suppressed DoS.  Finally, for times $t
\gg \tau$, diffraction is the dominant physical process.

Technically, the relevancy of diffraction follows from the fact that
for small energies $\epsilon \ll \hbar/\tau$ the scattering operator is
the dominant term in the Eilenberger equation of the normal region.
For any given real space point, $\br$, the scattering term involves an
average over all trajectories passing through $\br$. Apart from some
exponentially rare exceptions, the vast majority of these trajectories
will have a flight time of the order $t\approx t_{\rm f} \ll
\hbar /\epsilon$. For all these trajectories, the solution of the clean
Eilenberger equation is approximately given by $\sigma_2$, i.e. set by
the bulk value of the order parameter. This justifies the approximation
of the collision operator by ${\cal O}_{\rm coll} \approx -\hbar(2\tau)^{-1}
[\sigma_2,\;]$. Physically, ${\cal O}_{\rm coll}$ then acts like an
induced order parameter of strength $\approx \hbar/(2\tau)$.
This effective order parameter field in the N region describes the large angle
scattering of Andreev states off the induced proximity pairing field
amplitude. (Notice the difference between a pairing field amplitude,
$\langle \psi \psi \rangle$, and an order parameter, which is given by 
the coupling constant 
$\times \langle \psi \psi \rangle$).

The effect of ${\cal O}_{\rm coll}$ on the solution of the
equation depends on the length of the trajectory under consideration.
For trajectories with $t_0\ll \tau$, the clean solution
(\ref{solution_clean}) is close to the bulk form set by the presence
of the superconductor at the boundaries anyway, and the extra presence
of the diffraction operator is of no qualitative significance. For the
sparse set of exceptionally long trajectories with $t_0> 2\tau$,
however, the boundary condition is largely inessential and the
equation is governed by a competition between the energy and the
collision operator in the bulk N region. Inspection of the equation
shows that the solution will {\it approximately} be of standard 
BCS type
\begin{equation}
  \label{g_approx_small}
  g(\bxp)\Big|_{t_0(\bxp)> \tau} \stackrel{\epsilon < 
\hbar/(2\tau)}{\approx} 
{1\over \sqrt{(\hbar/2\tau)^2 - \epsilon^2}}\left(i\epsilon \sigma_3 
+ \frac{\hbar}{2\tau}
  \sigma_2\right). 
\end{equation}
Eq.~(\ref{g_approx_small}) leads to vanishing DoS for energies
$\epsilon < \hbar/(2\tau)$.

Summarizing we have formulated a primitive trial solution
$g_0(\bxp,\epsilon)$ to the Eilenberger equation that predicts a
gapped DoS. For trajectories $t_0(\bxp)\ll \tau$ (energies
$\epsilon\gg \hbar/\tau$), $g_0$ coincides with the diffractionless
solution (\ref{solution_clean}). Only for very long trajectories the
non-diffractive asymptotics is cut off by the collision operator,
leading to the trial solution of Eq.~(\ref{g_approx_small}).  We
emphasize that the simple BCS solution (\ref{g_approx_small}) is
specific to the case of purely $s$-wave scattering and becomes
unreliable in the vicinity of the gap, $\epsilon\approx
\hbar/(2\tau)$. In the next section we will explore the sensitivity of
the result to perturbative corrections around the trial solution. It
will turn out that the vanishing of the DoS below the gap is a
phenomenon robust to infinite order in perturbation theory.

\section{Field Theoretic Formulation}
\label{sec:sigma}

The previous discussion raises the question of whether the DoS gap
represents a generic feature of disordered billiards systems 
or whether it is an artefact of the BCS
type low energy trial solution derived above. In principle, one may
address this issue by using the trial solution as a starting point for
an infinite order perturbative expansion of the Eilenberger equation.
(Given that we are looking for residual corrections to a {\it
  vanishing} DoS, any conclusive perturbative approach has to be
driven to infinite order.)  In practice, however, such a type of
perturbative expansion is neither easy to formulate nor does it
provide much physical insight.  Given this situation we find it more
convenient to follow a different approach.  This alternative route,
which will be based on a variational principle, has three advantages:
(i) the variational approach will provide us with a comparatively
transparent verification of the robustness of the gap, (ii) it draws
upon analogies between the current problem and more general recent
developments\cite{KhM2,A3S:Nph,altland99} regarding the dynamics of
chaotic mesoscopic systems, (iii) it will provide the theoretical
background for the  subsequent analysis
of clean chaotic SN systems.

What is the meaning of the phrase `variational principle' in the
present context?  Within the framework of normal systems, Khmelnitskii
and Muzykhantskii have found\cite{KhM2} a general scheme whereby
quasiclassical kinetic equations are obtained through a variational
analysis of certain functional integrals of nonlinear $\sigma$-model
type. Below we will introduce the (straightforward) extension of this
scheme to the case of superconductor problems. Thereafter the
functional integral formalism will be applied to an analysis of the
DoS. By integrating out fluctuations around the trial solution
discussed in the previous section, we will find that the gap survives.

To discuss the variational principle, we need to introduce some field
theory oriented formalism which, at first sight, might appear to be
quite unrelated to the previously discussed material: Consider the
functional integral
$$
Z[\hat J] \equiv \int {\cal D}T e^{-S[T]},
$$
where the action $S[T] = S_{\rm reg}[T] + S_{\rm coll}[T]$ with 
\begin{eqnarray}
\label{ball_action}
&&S_{0}[T]=-\pi \rho  \int d\bxp {\,\rm
  str}\Big(\hbar T\sigma_3 {\cal L}T^{-1} + \nonumber\\
&&\hspace{2.0cm} + 
 i\sigma_3 (\epsilon + \hat \Delta)Q + \hat J\otimes
 E_{11}^{\sc bf}Q\Big),\nonumber\\
&&S_{\rm coll}[T] = {\pi \hbar \rho\over 4}
\int \frac{d\br d\bn d \bn'}{L^d(S_d)^2} w(\bn,\bn'){\,\rm str\,}\left(
Q(\br,\bn) Q(\br,\bn')\right).
\nonumber \\
\end{eqnarray}
Here $T(\bxp)$ and $Q(\bxp) = T(\bxp) \sigma_3 T(\bxp)^{-1}$ are four
dimensional supermatrix fields in phase space. As before, the
(unit-normalized) integration extends over a phase space shell of
constant energy $E_F$, and the inverse level spacing at $E_F$ is
denoted by $\rho$.  The rest of the notation is adopted to conventions
standard in the literature of the nonlinear
$\sigma$-model\cite{Efetbook}: The field $T$ takes values in the
(maximally Riemannian subset) of the graded coset space ${\rm
  GL}(2|2)/({\rm GL}(1|1) \otimes {\rm GL}(1|1))$. In an explicit
index representation, $ T(\bxp)=\{T_{a\alpha,b\beta}(\bxp)\}$ where
the indices $\alpha,\beta=1,2$ discriminate between the bosonic and
fermionic blocks of $T$ and the index $a,b=1,2$ is the {\sc ph} index.
'str' is the standard supertrace defined through ${\rm str}(A) = -
\sum_{a,\alpha} A_{a\alpha,a\alpha} (-)^{\alpha}$ (which differs by an
overall minus sign from Efetov's convention.)  Finally, the two
dimensional matrix field $\hat J=\{J_{ab}(\bxp)\}$ serves as a source
field to generate correlation functions, and $E_{11}^{\sc bf}$ is
defined as 
$(E_{11}^{\sc bf})_{ij}= \delta_{i1}\delta_{j1}$.

An action of the structure of (\ref{ball_action}) has first been
introduced in the context of normal systems with weak impurity
scattering\cite{KhM2}. (For an alternative derivation, see
Ref.\cite{A3S:Nph}.) For a normal system, $\hat \Delta = 0$ and the
action describes the correlated propagation of an retarded and an
advanced Green function in a ballistically disordered environment.
(The role of the indices $a,b$ then is to discriminate between the
advanced and the retarded sector of the theory.) Technically, the
functional integral $Z[\hat J]$ is of nonlinear $\sigma$-model type,
so that Eq.~(\ref{ball_action}) is commonly designated as the action
of the `ballistic $\sigma$-model'.

In the context of superconductivity, the model describes a {\it
  single} impurity averaged Gorkov Green function. More specifically,
\begin{eqnarray}
 G(\br,\bn) &=&
\frac{1}{\pi\nu_0}
\left. 
{\delta Z[\hat J] \over \delta \hat J(\br,\bn)}
\right|_{\hat
  J=0} \nonumber \\
&=& \langle Q_{11}(\br,\bn)\rangle_Q,
 \label{dZdJ}
\end{eqnarray}
where $\nu_0 = \rho/L^d$ and $G(\br,\bn)$ is the energy-integrated
Wigner transform of
the microscopic Gorkov Green function\cite{fn4}. Further, the
indices in $Q_{11}$ refer to {\sc bf} space, and $\langle \dots \rangle_Q$
denotes functional averaging with respect to the source free action
$S[Q]$.  From (\ref{dZdJ}) the DoS is obtained as
$$
\nu = {\nu_0\over 2}\int d\br d\bn {\,\rm Re\,}\langle {\rm
  tr\;}(Q_{11}(\br,\bn) 
\sigma_3)\rangle_Q.
$$
The derivation of both the effective action (\ref{ball_action}) and
the key relation (\ref{dZdJ}) is discussed in Appendix \ref{app:ball}.
We here merely notice that the construction of the model involves
essential elements of the quasiclassical approach. In fact, the
derivation essentially amounts to an implementation of the
quasiclassical construction steps -- disorder average, Wigner
transform, and integration over the modulus of the momentum $p$ -- into a
functional integral approach. (The additional graded structure of the
functional integral is physically of marginal relevance; supersymmetry
merely ensures unit normalization $Z[\hat J=0]$ of the source free
partition function. In fact, this aspect will not be of much
importance in our further analysis.)

What is the connection between the field integral approach and the
Eilenberger equation and, closely related, why  contemplate
a functional integral in a situation that is well described by
quasiclassics? To address these issues, we first notice that the
functional integral $\hat Z[\hat J]$ cannot be evaluated in closed
form. It is, however, a good candidate for a  stationary phase
approach. (This is because in the present application the coupling
constants appearing in the action are much larger than unity.) As usual with nonlinear $\sigma$-models,
the action $S[T]$ possesses not just one but rather an entire manifold
of degenerate saddle points (a fact that readily follows from the
observation that the action is invariant under transformations in
boson/fermion ({\sc bf}) space.) We start out by seeking for those
saddle points $\bar Q$ whose matrix structure is as simple as
possible. Given the {\sc bf} invariant structure of the action, one
anticipates that these saddle points will be proportional to unity in
{\sc bf} space, $\bar T = \{\bar T_{ab} \delta_{\alpha\beta}\}$.

Defining $\bar Q \equiv \bar T \sigma_3 \bar T^{-1}$, it is a
straightforward matter to show that the variation of the source free
action leads to 
\begin{eqnarray*}
&&  {\delta S[\bar T]\over \delta \bar T(\bxp)}=0\Rightarrow\\
&&\qquad \Rightarrow \hbar {\cal  L}\bar Q(\bxp)  - i[ \sigma_3(\epsilon +
\hat \Delta(\bxp)),
  \bar Q(\bxp)] = {\cal O}_{\rm coll.}[\bar Q],
\end{eqnarray*}
where 
\begin{eqnarray*}
{\cal O}_{\rm coll.}[\bar Q]=-\frac{\hbar}{2} \int\frac{d\bn'}{S_d}
w(\bn,\bn')[\bar Q(\br,\bn'), \bar Q(\br,\bn)].  
\end{eqnarray*}
The above equations state that
the ({\sc bf} diagonal) mean field configuration of the functional
integral are determined by the Eilenberger equation or, equivalently,
that $\bar Q=g$ is to be identified with the quasiclassical Green
function. This connection can be made explicit by
evaluating (\ref{dZdJ}) on the diagonal saddle point level which gives
$$
G(\br,\bn) \simeq e^{-S[\bar Q]} \bar Q (\br,\bn)
 = \bar Q(\br,\bn) = g(\br,\bn),
$$
where the second equality  is based on the fact that (due to supersymmetry)
the action $S[\bar Q]=0$.

Another way of interpreting this result is to say that the ballistic
$\sigma$-model implements a variational principle of quasiclassics, a
fact first realized in Ref.~\cite{KhM2} (with reference to normal systems.)
In the context of superconductivity this observation is of significant
practical relevance, the reason being that the variational action
(\ref{ball_action}) turns out to be a tool more flexibly applicable
than the Eilenberger equation itself.  More specifically,
\begin{itemize}
\item the field integral approach enables one to include quantum
  fluctuations around the mean field extremum. As was shown in
  Ref.\cite{altland00} (for the case of the dirty limit, $\Delta <
  \hbar /\tau$) such fluctuations may lead to a renormalization of
 the leading order
  quasiclassical results.
\item it is straightforward to extend the field integral formulation
  to the computation of the disorder average of more than one Green
  function. This enables one to analyze mesoscopic fluctuation
  effects\cite{altland00}.
\item in cases where the exact solution of the Eilenberger equation is
  unknown one may still organize the functional integral around an
  approximate mean field solution.  Provided that this configuration
  is not separated from the true Eilenberger solution by points of
  non-analyticity, no information is lost and a (sufficiently
  accurate) integration over fluctuations around the trial solution
  will produce correct results.
\end{itemize}
For our present application, the third of the above points is of
relevance.  To explore the DoS at low energies $\epsilon$ one may
perform a functional integration around an approximate solution of the
Eilenberger equation. In the next section this scheme will be applied
to the disordered billiards discussed in section \ref{sec:qc}.   
In section \ref{sec:clean_chaotic} we will then extend the discussion 
to systems with smooth potentials.

\section{Spectrum of an SN-System with Ballistic Disorder}
\label{sec:ballistic_box}

In this section we apply the strategy outlined at the end of the
previous section to explore the DoS of a disordered billiard (with
$s$-wave impurities) for 
energies $\epsilon \ll \hbar/\tau$.  For these low energies, it is
convenient to organize the functional integration around the reference 
configuration 
$$
Q_{\rm r} \equiv R \sigma_3 R^{-1} =  \sigma_2,
$$ 
where the rotation matrix is
$$
R \equiv \exp\left(i{\pi\over 4}\sigma_1\right).
$$
Comparison with the discussion of section \ref{sec:qc}
(Eqs.~(\ref{solution_clean}) and (\ref{g_approx_small})) indeed shows that
for $\epsilon \ll \hbar/\tau$, this configuration is close to the exact
Eilenberger solution (see below on what precisely is meant by `close'
in the present context.) Alternatively, one might organize the
integration around the trial solution (\ref{g_approx_small}). For
practical reasons, however, $Q_{\rm r} = \sigma_2$ is a more
convenient starting point.

To include fluctuations around $Q_{\rm r}$, we next introduce the 
parameterization
$$
Q\equiv R \tilde{Q} R^{-1},\qquad \tilde{Q} \equiv T\sigma_3 T^{-1},
$$
where $T(\bxp)$ is a four-dimensional supermatrix field on the
constant energy shell in phase space. Substitution of this ansatz 
 into the action of the functional integral leads to
 \begin{eqnarray}
\label{S_ball_dot}
&&S_{0}[T]=-\pi \rho  \int_{\rm N}
{d{\br} \over V}\int {d\bn\over S_d} {\rm
  str}\Big(\hbar v_F \bn \cdot T\sigma_3 \partial_{\br} T^{-1} - \nonumber\\
&&\hspace{2.0cm} - 
 i\epsilon \sigma_2 \tilde{Q}\Big),\\
&&S_{\rm coll}[T] = {\pi  \hbar\rho\over 4 \tau}
\int_{\rm N} {d\br\over V} \int\frac{d\bn d\bn'}{S_d^2}
{\,\rm str\,}\left(
\tilde{Q}(\br,\bn) \tilde{Q}(\br,\bn')\right).\nonumber
\end{eqnarray}
The functional expectation value associated with the Green function
assumes the form
$$
G(\bxp) = \pi \nu_0 \langle (R \tilde{Q}R^{-1})_{11}(\bxp)\rangle_{\tilde{Q}},
$$
which leads to 
$$
\nu = -{\nu_0 \over 2} \int d\bxp {\,\rm Re\,}\langle {\rm
  str}(\tilde{Q}_{11}(\bxp) \sigma_2) \rangle_{\tilde{Q}},
$$
for the DoS in the normal region.  In (\ref{S_ball_dot}) we have
used that for the problem under consideration the Liouville operator
${\cal L}$ assumes the form $v_F \bn\cdot \partial_{\br}$.  The symbol
$\int_{\rm N} d{\br}$ indicates that the integration is restricted to
the N-region of the system (with volume $V$). At the boundary to the
superconductor, we have $T=\openone$, which implies that $Q$
becomes tightly locked to its bulk superconductor value, $\sigma_2$.  
Comparison
with the general expression (\ref{ball_action}) shows that (i) due to
transformation by $R$ the Pauli matrix multiplying the energy argument
 has rotated to the superconductor reference
point ($\sigma_3 \to -\sigma_2$), and that (ii) in 
Eq.~(\ref{S_ball_dot}) the appearance of the
order parameter is implicit, namely through the boundary condition
imposed at the SN interface.

To prepare the integration over fluctuations, we next introduce the
standard\cite{Efetbook} system of coordinates $T=1+W/2$, where
the matrices $W$ are defined through
$$
W=\left(\matrix{&B\cr \bar B&}\right),
$$
the block decomposition is in {\sc ph} space and the
$2\times2$ supermatrices $B$ are subject to the (convergence ensuring) 
relation
$$
\bar B =  \sigma_3^{\sc bf} B^\dagger.
$$
We substitute this representation into the action
(\ref{S_ball_dot}) and expand formally according to
$$
S[T] \to S[B,\bar B] \equiv \sum_{n=1}^\infty S^{(n)}[B,\bar B],
$$
where $S^{(n)}[B,\bar B]$ is the contribution of $n$th order in
$B,\bar B$ to (\ref{S_ball_dot}). Similarly, the DoS functional
expectation value can be represented as
\begin{equation}
  \label{DoS_exp_val}
\nu =  {\nu_0\over 2}\int_{\rm N} d\bxp
\sum_{n=0}^{\infty}\langle
D^{(n)}[B,\bar B]\rangle_{B,\bar B},
\end{equation}
where the operators $D^{(n)}[B,\bar B]$ are defined through the series 
expansion
$$
{\rm
  str}(\tilde{Q}_{11}(\bxp)\sigma_2) = \sum_{n=0}^{\infty} D^{(2n+1)}[B,\bar
B],
$$
and we have noticed that only contributions of odd power in $B$
appear.  
To prepare the perturbative 
integration over the $B$ degrees of freedom we next consider the 
second order operator $S^{(2)}[B,\bar B]$ of the action. It is a
straightforward matter to verify that the explicit form of 
$S^{(2)}[B,\bar B]$ is given by 
\end{multicols}
\widetext
\begin{eqnarray*}
S^{(2)}[B,\bar B]= 
\frac{\hbar\pi \rho}{2}  \int_{\rm N}
{d{\br} \over V}\int {d\bn\over 2\pi}{d\bn'\over S_d^2} \, {\rm
  str}\left[B(\br,\bn)\left(\left( -v_F \bn 
\cdot \partial_{\br}+{1\over    \tau}\right)
\delta(\bn-\bn')- {1\over \tau}\right)\bar B(\br,\bn')\right].
\end{eqnarray*}
\begin{multicols}{2}
\narrowtext\noindent
The operator 
$$
\Pi^{-1}\equiv \frac{\hbar\pi\rho}{2}
\left(\left( -v_F \bn \cdot \partial_{\br}+{1\over
    \tau}\right)
\delta(\bn-\bn')- {1\over \tau}\right)
$$
governing the quadratic action has two key structural features
(which, as will be discussed in the next section, generalize to
different system categories): it is (a) real and (b) possesses a
spectrum that is gapped against zero. Whereas the reality of
$\Pi^{-1}$ is manifest, the
presence of a  gap follows from the fact that Dirichlet boundary
conditions $B=0$ at the interface to the superconductor are
imposed. This implies that the phase space zero mode $B(\bx)={\rm
  const.}$, the only candidate for a zero energy eigenmode, does not
belong to the representation space of $\Pi^{-1}$. It is
straightforward to verify that all eigenmodes with a finite variation
have eigenvalues of at least ${\cal O}(\hbar/\tau)$. 

The features (a) and (b) imply that the operator $\Pi^{-1}$ possesses
a real and non-singular inverse $\Pi(\bxp,\bxp')$. As we will discuss
in a moment, these properties suffice to essentially determine the
structure of the low energy DoS.

To perturbatively integrate out the $B$'s, we employ the supersymmetry 
implementation\cite{Efetbook,altland99} of Wick's theorem: the basic 
Gaussian pair contraction of two matrices $B$ and $\bar B$ is
determined by the rules
\begin{eqnarray*}
&&  \left\langle {\rm str\,}(B(\bxp) A \bar B(\bxp')
    A')\right\rangle_0 = 
\Pi(\bxp',\bxp) {\,\rm str\,}(A){\,\rm
    str\,}(A'),\\
&&  \left\langle {\rm str\,}(B(\bxp) A){\,\rm str\,}(\bar B(\bxp')
    A')\right\rangle_0 =\Pi(\bxp',\bxp) {\,\rm str\,}(AA'),
\end{eqnarray*}
where $A$ and $A'$ are arbitrary two-dimensional supermatrices, and
$\langle \dots \rangle_0 \equiv \int {\cal D}(B,\bar B)
\exp(-S^{(2)}[B,\bar B]) (\dots)$ denotes averaging with respect to
the Gaussian action. The Gaussian average 
$$
\langle {\cal O}_1^{(n_1)}[B,\bar B]  {\cal O}_2^{(n_2)}[B,\bar
B]\dots \rangle_0
$$
of an arbitrary constellation of $n_k$th order operators ${\cal
  O}_k^{(n_k)}$ can then, at least in principle, be computed as the
sum of all possible total pair contractions.

In general, the infinite series emanating from an expansion of the
exponentiated action is far too complex to be perturbatively brought
under control. With the current problem, however, the situation is
different. It turns out that the series expansion of both the
operators entering the action and the preexponential terms are
sufficiently highly structured to make possible some categoric
(infinite order in perturbation theory) statements.  To be more specific,
notice that the operators appearing in the action fall into two
different categories:
\begin{itemize}
\item operators $O_{\rm e}[T] = \sum_{n}{\cal
    O}_{\rm e}^{(2n)}[B,\bar B]$ whose expansion contains only {\it even}
  powers of $B$ and which are real (in the sense that the
  operator content multiplying the $B$'s is real). The two representatives
  of this category are the kinetic operator, $\sim {\rm
    str}( \bn \cdot T\sigma_3 \partial_{\br} T^{-1})$, and the
  collision operator $\sim {\,\rm str\,}\left( \langle
    \tilde{Q}\rangle_{\bn} \langle \tilde{Q}\rangle_{\bn}\right)$
\item operators of the type $i O_{\rm o}[T] =  \sum_{n}{\cal
    O}_{\rm o}^{(2n+1)}[B,\bar B]$ where $O_{\rm
    o}$ is again real but contains only {\rm odd} powers of $B$. This
  class of operators is represented by the energy term, $\sim 
  {\rm str}(\sigma_2 \tilde{Q})$. 
\end{itemize}
Similarly we notice that the preexponential expansion
(\ref{DoS_exp_val}) belongs to the class of purely odd
operators. However, contrary to the operators appearing in the action, 
the preexponential expansion is purely {\it imaginary} (the $i$'s
appearing in the definition of the Pauli matrix $\sigma_2$.) 
Finally, we repeat that the `propagator' $\Pi$ is real.

With all these things in place we now investigate the perturbative
representation of the DoS expectation value. Consider the contribution 
of any particular  operator $D^{(2n+1)}[B,\bar B]$ to the
expansion (\ref{DoS_exp_val}). In order to obtain a non-vanishing
contraction, we need to bring down a complementary odd operator from the
action. The most general odd operator that can be constructed from the 
exponentiated action is of the form
$$
 i^k O_{\rm o}^{(2n_1+1)}\dots O_{\rm o}^{(2n_k+1)} O_{\rm
  e}^{(2m_1)} \dots O_{\rm e}^{(2m_l)},
$$
where $l$ is arbitrary but $k$ must be odd. The total contraction 
\begin{eqnarray*}
&&  \left\langle D^{(2n+1)} O_{\rm o}^{(2n_1+1)}\dots O_{\rm
    o}^{(2n_k+1)} O_{\rm 
  e}^{(2m_1)} \dots O_{\rm e}^{(2m_l)} \right\rangle = \\
&&\hspace{4.0cm}=i \times {\, \rm
  real}  
\end{eqnarray*}
is then purely imaginary, which implies that {\em no contributions} to
the DoS, Eq.~(\ref{DoS_exp_val}), obtain. 
Summarizing, we have found that the mere
complex structure of the operators coupling to the DoS entails the
absence of perturbative DoS corrections. Before moving on, let us make 
a few general remarks on the structure of the argument:
\begin{itemize}
\item The construction formulated above explicitly relies on the fact
  that the energy operator in the rotated representation is purely
  imaginary. At first sight it looks like this makes the derivation
  explicitly gauge dependent. Indeed, for order parameter phases
  $\phi\not=0$ the rotation to the superconductor configuration
  introduces operators into the action that are neither purely real
  nor imaginary. However, it is straightforward to verify that (a) the
  phase couples only to the class of odd operators and (b) that any
  non-vanishing Wick contraction contains an equal number of phases
  $\exp(i\phi)$ and $\exp(-i\phi)$. This ensures the gauge invariance
  of the argument.
\item We emphasize that, in spite of the fact that the construction is
  based on the zero energy Eilenberger solution $g=\sigma_2$, it is by
  no means equivalent to a mere perturbative expansion of the DoS around zero
  energy. In fact we might have used any other low energy approximate
  solution $Q_{\rm r}$ -- in particular, the trial solution,
  Eq.~(\ref{g_approx_small}), may also have served as a reference
  configuration. As long as $Q_{\rm r}$ is
  not separated from the 'true' Eilenberger solution $\bar Q$ by
  non-analyticities, $\bar Q$ can be reached from $Q_{\rm r}$ by a
  perturbative expansion in $W$, and the conclusion of a vanishing DoS
  remains valid. The absence of non-analyticities for small energies
  is in turn guaranteed by the presence of the diffraction operator
  ${\cal O}_{\rm coll}$.  As follows from the structure of the
  effective action, $ {\cal O}_{\rm coll}$ causes the induction of a
  stable BCS gap $ \sim \hbar/(2\tau)$ on  {\it all} 
  classical trajectories  in the normal region. For subgap energies,
  $\epsilon\ll \hbar/\tau$, the mean field equations of the theory are
  regular and the perturbative scheme works. (The same is true for the
  regime of large energies $\epsilon \gg \hbar/\tau$ where the
  Bohr-Sommerfeld DoS of the non-diffractive system is retrieved.)  In
  the vicinity of the gap region, however, the theory is fundamentally
  unstable and can not be used to produce reliable results.
\item The stabilization of the gap by a non-analyticity in the
  functional integral can explicitly be tested for the particular case
  of {\it diffusive} SN systems. In the diffusive regime, the
  Eilenberger equation transmutes to the Usadel equation which, at
  least for a number of relevant geometries, can be solved in closed
  form (cf. the discussion of section \ref{sec:qualitative}.) A field
  integral analysis of fluctuations the Usadel mean field
  solutions\cite{altland00} then shows that no perturbative corrections to
  the zero subgap DoS exist. The responsible mechanism is perfectly
  analogous to the one discussed above.
\end{itemize}

We remark also, while we have demonstrated the absence of perturbative
corrections to the zero subgap DoS, this conclusion does not rule out
the possibility of contributions to the subgap DoS of a
nonperturbative nature. For diffusive systems, such contributions to
the subgap DoS do exist, as pointed out in Ref.~\cite{altland00} and
later investigated in detail\cite{skvortsov}, and are associated with
a rare class of states, known as `tail' states. These states are
poorly contacted with the superconductor and can exist below the gap
to give rise to contributions to the DoS that are, for diffusive
systems, exponentially small in the dimensionless conductance.  In
terms of the field theory, such states correspond to configurations of
the $Q$-matrix that are perturbatively inaccessible from both the
conventional Eilenberger saddle-point and the metallic saddle-point of
the theory. (More precisely, such states can be associated with saddle
points that break causality in the fermion-fermion sector of the
theory.)  The study of tail states in the ballistic regime, while an
interesting and unexplored topic, is beyond the scope of the present
work.

Finally, to prepare the analysis of soft scattering systems, we
introduce a bridge between quasiclassics/field theory and the
qualitative trajectory oriented pictures discussed in section
\ref{sec:qualitative}. In order to keep the discussion simple, it is
convenient to consider a situation where the non-diffractive
Eilenberger solution is close to the metallic reference point (large
energies) and can be interpreted as the quantum mechanical amplitude
of pair propagation of ordinary electrons and holes. The extension to
strongly proximity influenced regimes is technically straightforward
but less easy to visualize.

The most basic constituent of the field theory discussed above is the
propagator $\Pi$, i.e. the (inverse of the) kernel of
the quadratic action.  Expanding the action around the metallic
reference configuration $Q=\sigma_3$ it is straightforward to show
that in the diffractionless limit, the kernel is given by 
\begin{equation}
  \label{propagator}
\Pi^{-1} = \frac{\pi\rho}{2}\left(-\hbar{\cal L}   - 2 i \epsilon\right).  
\end{equation}
This means that the propagator of the theory in phase space,
$\Pi(\bxp,\bxp')$ is just the $\epsilon$-Fourier transform of the
classical time evolution operator,
\begin{eqnarray*}
&&\Pi(\bxp,\bxp') = {2\over \pi\hbar \rho} 
\int_0^\infty dt e^{2i\epsilon t/\hbar}
\left(e^{{\cal L} t}\right)(\bxp,\bxp')=\\
&&\hspace{3.0cm}=
{2\over \pi\hbar\rho} \int_0^\infty dt e^{2i\epsilon t/\hbar}
\delta(\bxp-\bxp'(t)).
\end{eqnarray*}
By expanding the functional integral in terms of the $B$ field,
one may now describe the effect of Andreev
scattering and/or quantum diffraction in terms  of
perturbative superpositions of classical propagators. For example, 
expanding the functional to lowest order around
the quadratic kernel the first correction to the metallic DoS is found 
to be 
$$
\delta \nu =-
\pi\rho \Delta^2\nu_0 \,{\rm Im}\,\frac{\partial}{\partial\epsilon}
\int_S d\bxp_1 d\bxp_2\, \Pi(\bxp_1,\bxp_2),
$$
where the $\Pi$ describes the propagation along the classical
trajectory of an Andreev bound state (Fig.~\ref{fig:hikami},
right), and $\int_S$ denotes an integration over the superconductor
interface.

\begin{figure}[h]
  \centerline{\epsfxsize=2.5in
  \epsfbox{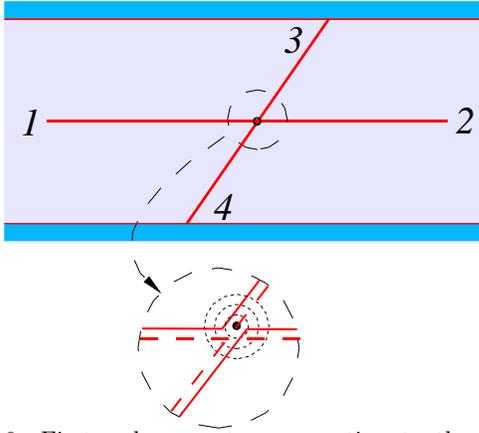}}
  \caption{First order quantum correction to the classical propagator
    in a SN-system with hard impurity scattering. Inset: solid line:
    propagation amplitude of an electron, dashed line: amplitude of
    the Andreev reflected hole state, dot: $s$-wave impurity
    scatterer.}
  \label{fig:diffraction_hard}
\end{figure}
In the same manner one verifies that, at next order in an expansion of
the functional integral in the generators $B$,  a quantum 
correction to the classical propagation of the type 
\begin{eqnarray}
&&\delta \Pi(\bxp_1,\bxp_2) = -2\frac{\hbar\Delta^2(\pi\rho)^3}{\tau}
\int_N \frac{d\br}{V}\int \frac{d\bn'd\bn}{S_d^2} \int_S
d\bxp_3 d\bxp_4 \times\nonumber\\
&&\times 
\Pi(\br,\bn;\bxp_1) \Pi(\bxp_3;\br,\bn) 
\Pi(\br,\bn';\bxp_4)
\Pi(\bxp_2;\br,\bn')+\dots ,
\label{deltaPi_hard}
\end{eqnarray}
appears. (Here the ellipses denote contributions of similar architecture but
somewhat different structure in phase space.)

The integration $\int_S$ extends over (the phase space points on) the
superconductor interface.  A cartoon of this process is depicted in
Fig.~\ref{fig:diffraction_hard}, where the inset represents a
visualization of amplitudes involved in the diffractive $s$-wave
scattering.  We notice that $\delta \Pi$ represents the quantitative
formulation (for a hard scattering ballistic environment) of the
diffraction corrections discussed qualitatively in section
\ref{sec:qualitative}. Through such processes, an exceptionally long
classical trajectory (quantitatively described by the propagator
$\Pi(\bxp,\bxp')$) establishes contact with the superconductor on time
scales $\tau$. Substitution of the modified propagator into the
formula for $\delta\nu$ gives the first quantum
correction to the DoS. 

Technically, the quantum contribution to $\Pi$ can be interpreted as
some `self energy' correction of strength $\sim \hbar/\tau$.  On small
energy scales $\epsilon < \hbar/\tau$ the action of the diffraction
operator is no longer perturbatively accessible. Nonetheless, the
concept of a low order expansion around the limit of classical
propagation remains a useful concept: In cases where the mechanism of
quantum diffraction is less trivial than mere $s$-wave scattering, a
perturbative analysis can be used to determine the characteristic
energy scale beyond which the quantum propagation becomes dominated by 
diffractive mixing between classically isolated regions in phase
space. The general structure of the theory implies that this time
scale sets the width of the gap. We will come back
to this point in the next section, when the soft scattering problem is
discussed.

\section{Clean Chaotic Systems}
\label{sec:clean_chaotic}

In this section we turn our attention to the proximity effect in clean
chaotic systems. As before we focus on billiard type systems in which
the average time of flight $t_{\rm f}$ represents the shortest time
scale in the problem. We assume that the scattering potential is
smooth. (In billiards, i.e. systems where the characteristic length
scale of the scattering potential $a$ coincides with the system size,
'smooth' means $a \gg \lambda_F$. For a definition applicable under
more general conditions, see\cite{AlLa1}.)

The
traditional form of the Eilenberger equation, Eq.~(\ref{eilenberger}),
applies to billiard type systems subject to weak impurity scattering.
Its naive generalization to generic clean systems
characterized by a set of  phase space coordinates
$\{\bxp\}$ is given by
\begin{eqnarray}
  \label{eilenberger_gen}
&&  \left(\hbar{\cal  L}  - i[ \sigma_3 ( \epsilon + \hat \Delta(\bxp)),
  \;\;]\right)g(\bxp) = 0,
\end{eqnarray}
with the standard  non-linear normalization $g(\bxp)g(\bxp)=\openone$.
This equation differs from   (\ref{eilenberger}) by (i) the absence of 
an impurity 
scattering operator ${\cal O}_{\rm coll}$ and (ii) the generalization
from $v_F {\bf n} \cdot \partial_{\bf r}$ to an 
arbitrary Liouville operator.

A moment's thought, however, shows that Eq.~(\ref{eilenberger_gen})
poses a serious problem: as discussed in section \ref{sec:qc},
the diffractionless Eilenberger equation reproduces the BS result for
the DoS. (No specific assumptions on the structure of the Liouville
operator were made in that argument.) Relying on
(\ref{eilenberger_gen}) one would thus have to conclude that the
proximity DoS of clean chaotic systems is not gapped, i.e. that the
DoS gap must be a peculiarity of disordered systems.  This hypothesis,
however, can be dismissed on the basis of a simple argument: being a
positive definite quantity, the DoS of disordered systems can only
vanish on average, if it vanishes for each individual realization.
However, a given realization of a disordered system is nothing more
than a Hamiltonian chaotic system. From this argument one expects that
the proximity gap is a generic signature of chaotic dynamics (save,
perhaps, for chaotic systems -- unknown to us -- that do not fall into
the universality class of standard Hamiltonian systems with chaotic potential
scattering.)

The resolution of the paradox is that for clean chaotic systems the
derivation of (\ref{eilenberger_gen}) is {\em ill-founded}. Indeed, the
derivation of the Eilenberger equation is based on the presumption
that the quasiclassical Green functions varies slowly on the scale of
the Fermi wave length. For disordered systems, the ensemble averaging
over realizations makes sure that this requirement is met. With
chaotic systems, however, the situation is different. The phase space
eigenfunctions of the chaotic Liouville operator are singular and
fluctuate on all length scales. This invalidates any semiclassical
gradient expansion, {\it unless} a smoothing regularization 
mechanism is active. 

Before discussing the issue of regularization further, we switch
back to the language of the field integral formulation.  The
variational action corresponding to the clean Eilenberger equation is
given by $S_0[T]$ of Eq.~(\ref{ball_action}). After the rotation $T\to
RT$, $R=\exp(i\pi \sigma_1/4)$ to the superconductor reference point
($\bar Q= \sigma_2$), the action simplifies to 
\begin{eqnarray}
  \label{S_chaotic_dot}
S_{0}[T]=-\pi \rho \int d\bxp {\rm str}\Big(\hbar
T\sigma_3 {\cal L}T^{-1} - i\epsilon \sigma_2 Q\Big)
\end{eqnarray}
where  coupling to the superconductor is implicit, through the
boundary condition $T=\openone$ at the SN-interface. 

Within the context of the general ballistic $\sigma$-model, the
necessity to regularize the effective action $S_0$ has been pointed
out earlier\cite{A3S:Nph,zirnbauer99,altland99}. More specifically, it
has been noticed that the phase space singularity alluded to above
invalidates the gradient expansion on which the derivation of the
effective action $S_0$ is based. To give the effective action some
meaning, a regularizing contribution $S_{\rm reg}$ has to be added.
Frustratingly, however, no generally accepted consensus regarding the
concrete implementation of a regularization operator has been reached
as yet. In fact, various different schemes have been proposed in the
literature: in the early work \cite{A3S:Nph} it was suggested to
regularize the action by adding a second order differential operator
of infinitesimal coupling strength.  In Ref.  \cite{AlLa1} it was
argued that a Hamiltonian chaotic system will {\it self}-generate some
regularization if only the mechanism of quantum dispersion is
consistently taken into account.  In lack of a fully microscopic
description of diffraction within a quasiclassical formalism, it was
suggested to model {\it phenomenologically} the effect of the quantum
spreading of minimal wave packets.  This was done by adding the
diffusion type second order operator
$$
S_{\rm reg}[Q] = -{\hbar\pi \nu\over 4\tau_{\rm r}} \int d\bx {\, \rm
  str\,}(\partial_{\bf n} Q \partial_{\bf n} Q)
$$
to the effective action, where the value of the characteristic
scattering time $\tau_{\rm r}= \lambda^2 v_F/\lambda_F$ and $\lambda$
is the largest Liapunuov exponent of the system (i.e. unlike in
Ref.\cite{A3S:Nph}, the coupling strength of the regulator was kept
finite.) Physically, the operator $S_{\rm reg}[Q]$ describes the
effect of some {\it soft} (low momentum transfer) scattering off a
stochastic potential; this may be seen readily by comparison with the
collision term $S_{\rm coll}$ in the action (\ref{ball_action}), in
the case where $w(\bn-\bn')$ is strongly peaked around $\bn-\bn'\sim 0$. 
We re-emphasize that this potential is
fictitious and only serves to model the spreading inherent to the
propagation of any quantum minimal wave package. In contrast, Ref.
\cite{zirnbauer99} argued that a chaotic system will {\it not}
auto-regularize itself and that some external smoothing is needed. It
was suggested to effect this regularization by adding to the action an
elliptic second order differential operator coupled to some weak
stochastic background field.  Again in contradiction to Ref.
\cite{AlLa1}, it was conjectured that a coupling strength vanishing
in the classical limit would be sufficient to generate regularization.

Although no general consensus regarding the practical realization  
has been reached, it is clear that 
the ballistic $\sigma$-model needs regularization 
through some
positive second order differential operator (or any other suitable
scheme.) In view of  this situation, we
here take a pragmatic point of view and add to  
the action $S_0$ the contribution
\begin{equation}
  \label{S_reg}
S_{\rm reg} \equiv -{\hbar\pi\rho D_{ij} \over 4}\int d\bxp {\, \rm
  str\,}(\partial_i Q \partial_jQ),  
\end{equation}
where $D_{ij}$ is a positive matrix whose characteristics will be
left unspecified for the moment.

Before moving on it is perhaps worthwhile to re-emphasize that this
regularization problem is in no way particular to the field theoretical
approach. It rather affects {\it all} semiclassical expansion schemes,
including those underlying the derivation of the Eilenberger equation.
In applications involving disorder, the averaging over a continuous
set of potential variables effectively smoothens out all
singularities, and the derivation of effective low energy theories is
not met with difficulties. For individual representatives of chaotic
systems, however, the problem is virulent and regularization must be
an integral part of any quasiclassical approach. We re-emphasize that
the actual mechanism by which regularization is effected `in reality'
is not well understood at present. 

We now turn to the discussion of the effective action $S=S_0 + S_{\rm
  reg}$ determined by Eqs.  (\ref{S_chaotic_dot}) and (\ref{S_reg}),
respectively.  Comparison with (\ref{S_ball_dot}) shows that $S$ bears
much structural similarity with the action of the previously discussed
$s$-wave scattering system. The essential difference lies in the
structure of the scattering operators $S_{\rm coll}$ and $S_{\rm
  reg}$, respectively. Whilst $S_{\rm coll}$ describes large momentum
transfer scattering at a rate determined by $1/\tau$, the
diffractive scattering modelled by $S_{\rm reg}$ is soft and of a
strength determined by the matrix $D$.  Naively, one might conjecture
that this difference in the microscopic structure cannot be of much
concern and that $S_{\rm reg}$ acts qualitatively similar to $S_{\rm
  coll}$.  This anticipation is wrong. In contrast to $S_{\rm coll}$,
the operator $S_{\rm reg}$ leads to a formation of a gap {\it only}
for systems with underlying chaotic dynamics. Further, the width of
that gap depends only logarithmically on the coupling strength, $D$.

To understand the action of the diffraction operator we consider the
soft scattering analogue of the perturbative picture developed in the
end of the previous section. Expanding the action to second order in
the generators $B$, it is straightforward to verify that, in the
presence of soft diffraction,  the classical propagator
(\ref{propagator}) generalizes to 
$$
\Pi^{-1} = \frac{\pi\rho}{2}\left(-\hbar{\cal L} - 2i \epsilon -
 \hbar D_{ij}\partial_i \partial_j
\right).
$$
One further verifies that the first order quantum correction to the
propagation along a classical trajectory is given by 
\begin{eqnarray}
\label{weak_correction}
&&\delta \Pi(\bxp_1,\bxp_2) = 8 
(\pi\rho)^3\Delta^2  \hbar D_{ij}\int d\bxp \int_S
d\bxp_3 d\bxp_4 \times\\
&&\hspace{2.0cm}\times 
\partial_{\bxp_i }\Pi(\bxp,\bxp_1) \Pi(\bxp_3,\bxp) \Pi(\bxp,\bxp_4)
\partial_{\bxp_j}\Pi(\bxp_2,\bxp)\nonumber,   
\end{eqnarray}
(compare with expression (\ref{deltaPi_hard}) for $\delta\Pi$ in
the case of hard, $s$-wave scattering).
At first sight, the structure of this expression suggests that upon
sending the regulator strength to zero, the quantum corrections to
classical propagation vanish linearly with $D$. However, this
anticipation is too naive; it does not 
account for  potentially singular behavior of the $\bxp$-derivatives
acting on the propagator $\Pi$. 

To better understand the behavior of the quantum
correction for small $D$, we need to explore the structure of the regularized
propagator $\Pi$. Consider the time representation of the defining
equation of $\Pi$, 
\begin{equation}
  \label{Pi_eq}
\frac{\pi\rho}{2}
(d_t - {\cal L}_\bxp - D_{ij}\partial_{x_i} \partial_{x_j})
\Pi(\bxp,\bxp';t) = 
\delta(\bxp -\bxp')\delta(t),   
\end{equation}
where $t\ge 0$ and the subscript in ${\cal L}_\bxp$ indicates that
the Liouvillian acts on the argument $\bxp$. For notational
convenience, the factor of two multiplying the energy argument
$\epsilon$ has been included in the definition of the time Fourier
argument $t$. In the absence of
diffraction, $\Pi_{D=0}(\bxp,\bxp';t) = (2/\pi\rho) \delta(\bxp - \bxp'(t))$. 
For finite $D_{ij}$, $\Pi$ is no longer purely deterministic
but rather  describes a stochastic process of diffusion
centered around the classical trajectory $\bx'(t)$. For
 sufficiently small times, the argument $\bxp$ is close to $\bxp'(t)$
 and the equation for $\Pi$ can be linearized in $\by \equiv
 \bxp-\bxp'(t)$.  Introducing ${\cal L} \equiv I_{ij} \partial_{x_j} H
\partial_{x_i}$, where the matrix
$$
I=\left(\matrix{&\openone\cr-\openone&}\right)
$$
implements the symplectic structure of the phase space, the
linearized equation for $\Pi$ assumes the form
\begin{eqnarray*}
&&
\left(d_t - {\cal L}_{\bxp'(t)} - M_{ij} \partial_{y_i}y_j -
D_{ij}\partial_{y_i} \partial_{y_j}\right) \Pi(\bxp'(t) + \by,\bxp';t) \\ 
&&\qquad=\frac{2}{\pi\rho}\delta(\by)\delta(t),   
\end{eqnarray*}
with $M_{ij} \equiv I_{ik} \partial^2_{x_j x_k} H\big|_{\bxp =
  \bxp'(t)}$. Noticing that for any function $f(\bx'(t))$ that is not
explicitly dependent on $t$, we have $(d_t - {\cal L}_{\bxp'(t)})
f(\bx'(t))=0$, the differential equation can be rewritten as
\begin{eqnarray}
\label{fp_eq}
&&\left(d_t - M_{ij} y_j \partial_{y_i} -
D_{ij} \partial_{y_j}\partial_{y_i}\right) \Pi(\by;t) 
=\frac{2}{\pi\rho}\delta(\by)\delta(t),   
\end{eqnarray}
where the dependence on both $\bxp'(t)$ and the initial condition are
implicit, and the time derivative acts only on the explicit time
argument in $\Pi(\by;t)$. Eq.~(\ref{fp_eq}) is a differential equation
of Fokker-Planck type. Owing to the complex time dependence of the
matrix $M$ it is, of course, not analytically solvable. Nonetheless,
some qualitative statements regarding the structure of the solution
can be made: Eq.~(\ref{fp_eq}) is governed by a competition between
the diffusion operator $D_{ij}\partial_i \partial_j$, and the drift
operator $ M_{ij}  \partial_{y_i}y_j$.  By definition, the time
averaged eigenvalues of the `drift matrix' $M$ are the Liapunov
exponents of the chaotic scattering problem. On short time scales, the
dynamics of the equation is controlled by the diffusion term and we
expect the distribution $\Pi$ to describe a stochastic diffusive
spreading around the deterministic solution $\bxp'$. After a certain
threshold time, a crossover to a regime controlled by the drift
operator takes place. On larger time scales, the distribution $\Pi$
describes some exponential separation away from the trajectory
$\bxp'$, where the rate of separation is determined by the Liapunov
exponents.

To formulate these concepts in a somewhat more quantitative manner we
make the assumption (acceptable 
on short time scales) that
the matrix $M$ is approximately static. (In fact, all what we really need to
require is that the largest Liapunov exponent, $\lambda$, and the
associated direction in phase space are constant.) For a time
independent matrix $M$, Eq.~(\ref{fp_eq}) can be solved in closed
form\cite{risken}:
\begin{eqnarray*}
  \Pi(\by,t) = \frac{4}{\rho (2\pi)^{(2d+1)/2}}
 (\det\sigma(t))^{-1/2}
e^{-{1\over 2} \by^T \sigma(t)^{-1} \by },
\end{eqnarray*}
where the matrix-kernel $\sigma$ is given by
$$
\sigma(t) = 2\sum_{\alpha \beta} {e^{(\lambda_\alpha + \lambda_\beta)t}-1 
  \over \lambda_\alpha + \lambda_\beta} {\bf v}^\alpha  ({\bf
  v}^{\alpha T} {\bf D} {\bf v}^\beta) {\bf v}^{\beta T}.
$$

Here ${\bf v}^\alpha$ and $\lambda_\alpha$ are eigenvectors and
eigenvalues of the drift matrix $M$, respectively. The enumeration of
the eigenvalues is ordered such that $\lambda \equiv \lambda_0 >
\lambda_1 >\dots >\lambda_{2d-1}$. Finally, re-installing our original
set of variables, we arrive at
\begin{eqnarray*}
&&  \Pi(\bxp,\bxp',t) = \frac{4}{\rho(2\pi)^{(2d+1)/2}}
  (\det\sigma(t))^{-1/2}\times\\
&&\hspace{3.0cm}\times
e^{-{1\over 2} (\bxp-\bxp'(t))^T \sigma(t)^{-1} (\bxp- \bxp'(t))}
\end{eqnarray*}
for the short time approximation to the propagator $\Pi$.  From this
expression we recover the temporal picture discussed qualitatively
above: For times shorter than $\lambda^{-1}$, $\sigma \approx Dt$ and the
solution describes diffusion around the classical trajectory $\bxp'$.
On time scales larger than $\lambda^{-1}$,
some of the matrix elements of $\sigma$ begin
to diverge exponentially. Defining $x_0 = {\bf v}^{0 T}\bxp$ as the
projection of $\bxp$ onto the direction of strongest exponential
divergence, we find
\begin{equation}
  \label{Pi_liapunov}
\langle (x_0 - x'_0(t))^2 \rangle \stackrel{t\gg
  \lambda^{-1}}{\approx} {D_0 \over \lambda} e^{\lambda t},
\end{equation}
where $D_0   \equiv {\bf v}^{0T} D
{\bf v}^0$ denotes the relevant element of the diffusion matrix. Thus,
for $t>\lambda^{-1}$ the propagation is no longer confined to the vicinity of
$\bxp'(t)$ but rather branches out exponentially, at a rate determined by
the largest Liapunov exponent of the system. We emphasize that this type of
behavior is specific to chaotic systems. For an integrable system, no
globally positive exponents driving the separation process exist, and,
the weak diffusive smearing left aside, $\Pi$ does not
deviate from the predestined trajectory $\bxp'$.
 
We now turn back to our discussion of the quantum correction,
$\delta\Pi$ of Eq.~(\ref{weak_correction}). 
Specifically we wish to understand how much
time it takes before an exceptionally long classical trajectory
diffractively spreads out into the phase space regions strongly
coupled to the superconductor. The first thing to be observed is that
the prefactorial dependence of $\delta\Pi$ in 
Eq.~(\ref{weak_correction}) on the
diffusion matrix $D_{ij}$ is largely cancelled by the derivatives acting on
the propagators $\Pi$. Indeed, the diffraction correction at fixed
energy, $\epsilon$, is determined by the time Fourier transform of the
propagators $\Pi(\bxp,\bxp';t)$. Parametrically, the contribution of
short times $t<\lambda_\alpha^{-1}$ to the time Fourier integral of
$\partial_{\bxp} \Pi$ is given by
$$
\partial_{\bxp} \Pi \sim (\bxp -\bxp')\sigma^{-1} \Pi \sim
\sigma^{-1/2} \Pi \stackrel{t\ll \lambda_\alpha^{-1}}{\sim} (Dt)^{-1/2}\Pi,
$$
implying that the $D$-dependence introduced by the two derivatives
cancels against that of the prefactor.  From this rough estimate we
conclude that it will {\it not} be the naive timescale related to the
coupling strength of the
operator that sets the effective rate of diffraction, but rather some
other time scale.

To determine that scale we apply a modified version of an argument
first formulated by AL within the context of normal metal weak
localization: let us focus attention on the coordinate $x_0$ of
strongest exponential divergence. For the sake of concreteness, we
make the (inessential) assumption that $x_0$ represents some real space
coordinate. From the solution for $\Pi$ we find that for short time
scales $t<\lambda^{-1}$, $ \langle (x_0 - x'_0(t))^2 \rangle \approx
D_0t$. On the other hand, the dispersive quantum spreading of a
Gaussian minimal wave packet centered around the classical trajectory
would be estimated by $\langle (x_0 - x'_0(t))^2 \rangle \sim v_F
\lambda_F t$. We now follow the general philosophy of AL and postulate
that the mere function of the regularization operator is to
phenomenologically mimic this effect.  This leads to the estimate $D_0
\sim v_F \lambda_F$\cite{fn5}.

Having parametrically estimated the strength of the diffraction
operator, we next consider the quantum correction, $\delta\Pi$ of 
Eq.~(\ref{weak_correction}), in the `Liapunov regime' of time scales
$t>\lambda^{-1}$. To couple a formerly isolated exceptionally long
trajectory $\bxp':\bxp_1 \to \bxp_2$ to the background of average
trajectories and/or the superconductor interface, the propagators
$\Pi(\bxp_3,\bxp)$ and $\Pi(\bx,\bxp_4)$ `branching out' from the
point $\bxp$ of diffractive scattering, have to leave the immediate
phase space vicinity of the reference trajectory. A conservative
estimate of the threshold-deviation necessary to initiate that
coupling reads as
$$
\langle (x_0 - x'_0(t))^2 \rangle \approx {D_0 \over \lambda}
e^{\lambda t} \sim a^2,
$$
i.e. we demand that the spreading reaches a scale comparable
with the system geometry. Finally, estimating the Liapunov exponent as
$\lambda \sim v_F/a$, and using the above expression for $D_0$, we
arrive at
$$
t \approx {1\over \lambda} \ln\left({a\over \lambda_F}\right),
$$
which is the celebrated Ehrenfest time $t_{\rm E}$.

Summarizing we have found that for chaotic systems with smooth
potentials, $t_{\rm E}$ is the time scale at which quantum corrections
begin to significantly affect the classical propagation along
ballistic trajectories. On time scales larger than $t_{\rm E}$ any
formerly isolated long classical trajectory `merges' into the
background of surrounding trajectories of average flight time $t_{\rm f}$
(where $t_f$ is the escape time).
According to the general structure of the theory, the inverse of this
threshold time, $\hbar/t_{\rm E}$, sets the width of the DoS gap.

It is instructive to study this mechanism more explicitly on the basis
of the Eilenberger equation. This is done
in appendix \ref{sec:ee}, where we consider the prototypical
example of an infinitely long trajectory $\bxp'$ that, classically, would
never hit the superconductor interface. In the absence of diffraction, 
the Green function along that trajectory is metallic, 
$g(\bxp',\epsilon)=\sigma_3$, and gives a finite contribution to the 
DoS. For ${\cal O}_{\rm coll}$ weak but finite, the
behavior of $g$ close to $\bxp'$ is described by a Fokker-Planck
equation similar to the one discussed above. The solution of
this equation, detailed in Appendix \ref{sec:ee}, shows that for energies
$\epsilon \sim \hbar/t_E$, a crossover from $\sigma_3$ to the
off-diagonal configuration set by the surrounding superconductor
takes place. For still smaller energies the DoS vanishes. 

Before concluding, let us make some general critical remarks on the
status of the results obtained in this section. Unlike in the case of
hard point-like scatterers, the derivation and solution of 
quasiclassical equations (alias the mean field equation of the
ballistic $\sigma$-model) relied on various phenomenological assumptions:
\begin{itemize}
\item A number of assumptions regarding the nature of the Liapunov
  exponents of the system, and the solution of the linearized
  equations of motion have been made. Furthermore, some 
  relevant classical and quantum system parameters have been estimated 
  in a crude manner.
\item The all-important diffraction operator has not been
  microscopically derived but was introduced on phenomenological
  grounds.
\end{itemize}
We believe that the first of these points is not serious. The reason
is that the characteristic energy scale $\hbar/t_{\rm E}$ depends only
logarithmically on the modelling of the dynamics, and on the values
assigned to the parameters $D,\lambda,a$, etc. As long as a globally
positive Liapunov exponent exists, an energy scale of the order of
$\hbar/t_{\rm E}$ will be generated.  
The second point,
however, is more important. The absence of a consistent microscopic
derivation of a diffraction operator indeed represents a deficiency of
the current approach and, for that matter, of the general field theory
approach to quantum chaos. On the other hand, the presence of a
diffraction operator, of the coupling strength specified above, is
required for basic physical reasons. We therefore believe that, at
least phenomenologically, the current approach stands on solid ground.

Finally, we notice that, unlike in the case of the hard scattering
problem, it was not possible to bring the Eilenberger equation into a
form manifestly equivalent to a bulk BCS equation with induced order
parameter.  This implies that the current approach cannot rigorously establish
the existence of a hard gap edge for the soft scattering problem.
What we {\it could} see is that on energy scales of the order of the
inverse of the Ehrenfest time, a quantum mechanism suppressing the DoS
becomes operative. The fact that for lower energies the solutions of
the Eilenberger equation undergo a rotation towards the bulk
superconductor configuration (cf. Appendix \ref{sec:ee}) implies that
the BS approach qualitatively overestimates the low energy DoS of
chaotic systems.

\section{Summary}
\label{sec:discuss}

In this paper we have discussed various physical mechanisms relevant
to the phenomenology of the proximity effect in chaotic SN-systems.
Considering systems where chaotic dynamics is induced by a weak amount
of point-like impurities as a test case, we have demonstrated that (a)
the existence of a hard gap in the DoS is directly related to an
essential singularity in the quasiclassical equations for the Green
function and (b) that these singularities can be interpreted 
semiclassically as a non-perturbative accumulation of quantum interference
processes akin to weak localization corrections in normal metals.
These contributions are effectively summed up in the solutions of the
quasiclassical equations. Although closed solutions can be obtained
only in exceptional cases (for example in the excessively investigated case
of systems with diffusive dynamics) we have shown that the strict
vanishing of the DoS for low energies follows from 
combining elements of quasiclassics with the formalism
of the ballistic $\sigma$-model. While all perturbative corrections to 
the DoS vanish below the gap, we have remarked that exponentially
small contributions due to tail states are not ruled out by our
analysis, but remain beyond the scope of this work.

Turning to the far more complex case of {\it clean} chaotic systems we
have discussed various difficulties obstructing the construction of
quantitative analytical approaches. The perhaps most fundamental
problem is that the assumption of a spatially continuous Green
function underlying, for example, the quasiclassical formalism fails (due to
the singular distribution of the phase space eigenfunctions of the
Liouville operator.) In principle one may evade this difficulty by
averaging the theory over some set of external parameters. Structure
and strength of a minimum set of averaging parameters needed to ensure
continuity are not known at present. However, taking a pragmatic point
of view and {\it assuming} continuity of the Green function, it can be
demonstrated that for energy scales smaller than the inverse of the
Ehrenfest time, quantum diffraction mechanisms become
operative\cite{AlLa1}. For those low energies the DoS of SN systems is
suppressed below the value predicted by the semiclassical
Bohr-Sommerfeld approach. (Here it is assumed that the Ehrenfest time
is larger than the average classical escape time from the normal metal
region. The opposite regime has not been investigated so far.) Most
likely, the inverse of the Ehrenfest time marks the position of a hard
gap edge although, due to the problem stated above and other
difficulties related to the formulation of quantitative theories of
chaotic systems (see the text), this could not be rigorously shown. In
consideration of the difficulties mentioned above, and the fact that
established semiclassical techniques based on counting classical
trajectories fall short of producing a hard gap, a quantitative
theory of the proximity effect remains a challenge for future
approaches to chaotic sytems.

\acknowledgments

We are grateful to O. Agam, M. Janssen and Yu. Nazarov for useful discussion
during the course of this work and gratefully acknowledge the 
financial support of 
the `{\em Sonderforschungsbereich}' 237.
 
\appendix
\section{Derivation of the Ballistic Sigma-Model}
\label{app:ball}
In this section we derive the effective action (\ref{ball_action}) of
a weakly disordered sample in the presence of a superconducting order
parameter.  While, as discussed in the main text, an action of this
type for a normal metal was first proposed by Muzykantskii and
Khmelnitskii \cite{KhM2}, we follow here more closely the derivation
of Andreev {\em et al.} \cite{A3S:Nph}, with suitable modifications:
we allow for the presence of a superconducting order
parameter by introducing particle and hole components in the
functional integral.  The method relies on an averaging over a set
$N\gg 1$ of energy levels around the Fermi surface to construct the
$\sigma$-model in the absence of disorder.  The derivation employs
further procedures that are central also to the quasiclassical
approximation, including a Wigner-transform and integration over the
momentum coordinate normal to the energy shell.  The disorder is
treated as a perturbation to derive a term resembling a collision
integral in the action.  Since we present the generating functional
for a single retarded Green function, the role of the usual
advanced-retarded components in the field integral are taken over by
particle-hole components. For simplicity we restrict ourselves to the
case of broken time-reversal symmetry, although this simplification is
not essential and the generalization to other symmetry classes
is straightforward.

The starting point is the Gorkov Hamiltonian,
\begin{eqnarray*}
 \hat{H} = \hat{H}_0+V(\br)+\hat\Delta\sigma_3; \qquad 
\hat{H}_0 = \frac{\hat{\bf p}^2}{2m}.
\end{eqnarray*}
The generating function for the single-particle Green function
may be expressed as a field integral in the form
\begin{eqnarray*}
Z[\hat{J}] &=& 
\int D\Psi\times \\
&&\times \exp\left[-i\int d\br
\Psi^{\dagger}(\br)L
(\hat{G}^{-1}(\epsilon_0)+i\hat{J}E_{11}^{\sc bf})
\Psi(\br)\right]
\end{eqnarray*}
where $\hat{G}^{-1}(\epsilon_0) =\epsilon_0-\epsilon\sigma_3-\hat{H}$.
$\Psi$ represents a four-component superfield, containing
particle-hole and boson-fermion sectors.  The energy $\epsilon$
represents an energy difference between particle and hole sectors. 
The matrix $L$ is defined as
\begin{eqnarray*}
L= \left(\begin{array}{cc}
1 & \\
& \sigma_3^{\sc bf} \end{array}\right), 
\end{eqnarray*}
with the block decomposition being in particle-hole space. 
Two
dimensional matrix field $\hat J=\{J_{ab}(\bxp)\}$ serves as a source
field to generate correlation functions,  and
$E_{11}^{\sc bf}$ is
defined as 
$(E_{11}^{\sc bf})_{ij}= \delta_{i1}\delta_{j1}$.
Next we employ Gaussian averaging over energy levels 
according to
\begin{eqnarray*}
\langle \ldots \rangle_{\epsilon_0}=\int\frac{d\epsilon_0}{(2\pi
  N^2)^{1/2}}
\exp\left[-(\epsilon_0-\epsilon_F)^2/(2N^2)\right].
\end{eqnarray*}
This induces a nonlocal interaction
term in the generating functional:
\begin{eqnarray*}
\langle Z(\hat{J})\rangle_{\epsilon} &=& \int D\Psi
 \exp\left[-i\int d\br
\Psi^{\dagger}(\br)L
\times \right. \\&&\left. \times
(\hat{G}^{-1}(\epsilon_F)+i\hat{J}E_{11}^{\sc bf})\Psi(\br)
-S_{\rm int}[\Psi]\right], \\
S_{\rm int} &=& \frac{N^2}{2} \left(\int d\br \Psi^{\dagger}(\br)L\Psi(\br)
\right)^2.
\end{eqnarray*}
The interaction term is then decoupled with a matrix $Q(\br,\br')$ 
that is nonlocal in space: after integrating over $\Psi$ we find
\begin{eqnarray}
\langle Z(\hat{J})\rangle_{\epsilon_0}&=&\int DQ\exp\left[
\frac{1}{2}{\rm str}_{\br}Q^2
\right. \nonumber \\ &&\hspace{-1.5cm}\left.
-{\rm str}_{\br} \ln\left(
\hat{G}_0^{-1}(\hat{Q})-\sigma_3(\epsilon+\hat{\Delta})
-V+i\hat{J}E_{11}^{\sc bf}\right)\right],
\label{strln}
\end{eqnarray}
where $\hat{G}_0^{-1}=\epsilon_F-\hat{H}_0-iN Q$ and $Q$ obeys the
symmetry $Q^{\dagger}= L Q L$, where the dagger denotes a Hermitian
conjugate. The strategy now is to
simplify the action so as to isolate fluctuations associated with
energies much less than the Fermi energy.  To do so we search for a
value of $Q$ that minimizes the action and expand in fluctuations
around this saddle-point. In addition, we consider first the totally
clean case ($V(\br)=0$), while later we include the effects of ballistic
disorder.  The saddle-point equation now reads as
\begin{eqnarray*}
Q_0 G_0^{-1}(Q_0) = -iN,
\end{eqnarray*}
with the explicit solution
\begin{eqnarray*}
Q_0 = -i\frac{\epsilon_F-\hat{H}_0}{2N} 
+\left(1-\left(\frac{\epsilon_F-\hat{H}_0}
{2N}\right)^2\right)^{1/2}\sigma_3.
\end{eqnarray*}
To describe the low-energy, long-wavelength fluctuations of $Q$ around
the saddle-point\cite{fn6y}, we may write $Q=T Q_0 T^{-1}$, where
$T$ obeys the symmetry $T^{\dagger}= L T L$, and expand
the action to leading order in $1/N$: we find 
\begin{eqnarray}
S[Q] &=& -\frac{i}{N}{\rm str}\left[Q\left(\sigma_3(\epsilon
+\hat{\Delta})-i\hat{J}E_{11}^{\sc bf}
-T^{-1}[\hat{H}_0,T]\right)\right].
\nonumber \\&&
\label{leading}
\end{eqnarray}
At energies far below the Fermi energy, the leading contribution to
the above functional integral may be described by the semiclassical
approximation (note that the use of the term `semiclassical' has an
entirely different meaning from that used in the BS
approach; here it stands for an expansion of operator traces in powers
of $\hbar$). To define this approximation, we apply the Wigner
transformation
\begin{eqnarray*}
{\mathcal O}({\bf R},{\bf p}) = \int d(\delta{\bf r}) e^{i{\bf p}.\delta{\bf
 r}/\hbar}
{\mathcal O}\left({\bf R}+\frac{\delta{\bf r}}{2},
{\bf R}-\frac{\delta {\bf r}}{2}\right).
\end{eqnarray*}
According to the semiclassical approximation, the Wigner transform of
a product of operators may then be approximated as $({\cal O}_1{\cal
  O}_2)(\br,{\bf p}) = {\cal O}_1(\br,{\bf p}) {\cal O}_2(\br,{\bf p})
+ \hbar \{ {\cal O}_1(\br,{\bf p}), {\cal O}_2(\br,{\bf p})\} +
\dots$, where $\{\cdot,\cdot\}$ denotes the classical Poisson bracket.
Specifically, 
the semiclassical approximation of the commutator with the Hamiltonian
reads as
\begin{eqnarray*}
[\hat{H}_0,T]\to -i\hbar {\cal L}T=-i\hbar\{T,H_0\},
\end{eqnarray*}
where ${\cal L}$ is the Liouville operator.  Further, assuming that
(a) fluctuations transverse to the constant energy shell $H_0={\rm
  const.}$ are frozen out and (b) $Q_0\approx \sigma_3$ (see
Ref.\cite{altland99} for a discussion of these points) we may write
the semiclassical approximation of the matrices $Q$ as
\begin{eqnarray*}
Q(\bx) = T(\bx)\sigma_3T(\bx)^{-1},
\end{eqnarray*}
where $\bx$ is a shorthand notation for the $2d-1$-dimensional set of
phase space coordinates of the constant energy shell. For a billiard
type system with hard boundary scattering $\bx = (\br,\bn)$.  

Performing the integration over the coordinate transverse to the
energy shell (for constant $Q$), we obtain from Eq.~(\ref{leading})
the $S_0$ contribution to the action contained in
Eq.~(\ref{ball_action}). This part of the action is applicable to a
clean sample in the presence of energy averaging.

To include the effects of ballistic disorder, we may now reintroduce
the impurity potential $V(\br)$. Although for short range impurities
this part of the Hamiltonian may not be treated by the semiclassical
approximation, in the limit of weak disorder it may be treated as a
small perturbation in a similar manner to the symmetry breaking terms
$\hat{J}E_{11}^{\sc bf}$ and $\epsilon\sigma_3$ appearing in the
action (\ref{strln}). Expanding the logarithm to second order in $V$
and averaging over impurity configurations then leads to the collision
term $S_{\rm coll}$ in Eq.~(\ref{ball_action}).

It may be verified (see main text) that the saddle-point equation for
$\bar Q$ that
minimizes the effective action (\ref{ball_action}) is precisely the 
the quasiclassical Eilenberger equation, 
Eq.~(\ref{eilenberger}). 
The strength of the field theory approach however is that
it enables evaluation of the
single-particle, impurity-averaged Gorkov Green function even {\em beyond} the
quasiclassical approximation. Differentiation 
of the partition function with respect to the source term, $\hat J$, leads to
\begin{eqnarray}
G(\br,\bn) &=& \frac{1}{\pi\nu_0}\left.\frac{\partial
  Z[\hat{J}]}{\partial\hat{J}(\br,\bn)} 
\right|_{\hat{J}=0} \nonumber\\
&=&\int DQ e^{-S[Q]} Q_{11}(\br,\bn),
\label{gork}
\end{eqnarray}
where $G(\br,\bn)$ is the Wigner-transformed and energy-integrated
form of the Gorkov Green function, and the 11 indices refer to {\sc
  bf} space. 
While evaluation of the integral in
Eq.~(\ref{gork}) at the saddle-point level
reproduces the quasiclassical approximation for $G(\br,\bn)\simeq 
\bar{Q}(\br,\bn)$,
integration over the entire $Q$-manifold in principle reproduces the
exact expression for $G(\br,\bn)$.

\section{Appearance of the Ehrenfest time within the quasiclassical
  formulation}
\label{sec:ee}

In this appendix we explore the influence of the collision operator on
the solution of the Eilenberger equation in a chaotic environment.
Specifically, we consider the equation in the vicinity of a classical
trajectory $\bxp'$ of infinite length, i.e. a trajectory that never hits the
superconductor interface. This trajectory is meant to serve as a
caricature of the long trajectories that support the BS density of
states at low energies.

As long as the collision operator is not included, the quasiclassical
Green function on our trajectory assumes its metallic value,
$g(\bxp',\epsilon) = \sigma_3$. For coordinates $\bxp$ away from the
reference trajectory and/or in the presence of a finite collision term
we expect deviations from $\sigma_3$. To parameterize these
configurations we introduce the parameterization $g(\bxp) = e^{ W(\bxp)}
\sigma_3 e^{- W(\bxp)}$, where $W$ is a $2\times 2$ antiunitary matrix
generating rotations around $\sigma_3$. 
For all what follows it will be convenient to parameterize $ W$ as
$W =  W_+ + W_-$, where
$$
W_\pm = i w_{\pm} {1\over \sqrt 2} (\sigma_1 \mp i \sigma_2), \qquad
w_\pm \in {\cal R}.
$$
Anticipating that in the vicinity of $\bxp'$, $W$ will be small, we
next expand $g$ as $g = \sigma_3 + 2W \sigma_3 + \dots$ and linearize 
the Eilenberger equation 
$$
 \left(\hbar {\cal  L}  - i\epsilon [\sigma_3,
  \;\;]\right)g(\bxp) - \hbar D_{ij} \partial_i \left(g(\bxp)\partial_j
  g(\bxp)\right)= 0, 
$$
in $W$. The result reads as
$$
 \left(\pm \hbar{\cal  L}  + 2i\epsilon 
 + \hbar D_{ij} \partial^2_{ij}\right) w_\pm = 0.
$$
To make progress with this equation, it is again convenient to
transform to time space. Fourier transforming we obtain
$$
 \left(\pm {\cal  L}  - d_t
 + D_{ij} \partial^2_{ij}\right) w_\pm = 0,
$$
where, as in the main text, the factor of 2 multiplying the energy
argument
$\epsilon$ has been included in the definition of $t$. 
To give this equation some meaning, initial conditions have to be
specified. To this end we introduce a coordinate system where the
first of the $2f$ phase space coordinates $y,y_0,y_1,\dots,y_{2d-1}$
parameterizes our reference trajectory, whereas the remaining
coordinates span the complementary region.  We next assume than all
possible choices of initial conditions $w_{\pm}(\bx,t=0)\equiv
w_{\pm,0}(\bx)$ have the common feature that $w_{\pm,0}$ increases up to
$w_{\pm}= {\cal O}(1)$ as any of the coordinates $x_0,\dots,x_{2d-1}$
approaches the characteristic scale of the potential, $a$. What this
means is that for phase space points farther away from the exceptional
trajectory than $a$, the Green function is locked to the
superconductor order parameter. We notice that neither the specific
choice of the threshold value $x_i \approx a$ nor the rigorous
implementation of the initial condition sensitively affect the results
of our analysis; it suffices that $w_{\pm,0}(x_i\sim a)\sim 1$ holds
in some average sense. 

The formal solution of our initial condition problem now reads as
$$
w_\pm(\bx,t) = \int d\bxp' K_\pm (\bxp,\bar\bxp,t) w_{\pm,0}(\bar\bxp),
$$
where the kernel $K$ is defined through 
\begin{eqnarray*}
&& \left(\pm {\cal  L}  - d_t
 + D_{ij} \partial^2_{ij}\right) K_\pm(\bxp,\bar \bxp,t) = \delta(t)
\delta(\bxp - \bar \bxp).
\end{eqnarray*}
This  equation is equivalent to Eq.~(\ref{Pi_eq}) for the propagator
$\Pi$ and we can directly transcribe the solution. Substituting 
\begin{eqnarray*}
K_\pm(\bxp-\bar\bxp,t) &=& \mp(2\pi)^{-(2d-1)/2} 
(\det \sigma_\pm)^{-1/2} \times \\
&&\times e^{- {1\over 2}
  (\bxp-\bar \bxp(t))\sigma_\pm^{-1}(\bxp-\bar \bxp(t))}
\end{eqnarray*}
with 
$$
\sigma_\pm(t) = \pm 2\sum_{\alpha \beta} \frac{
e^{\pm(\lambda_\alpha + \lambda_\beta)t}-1} 
{\lambda_\alpha + \lambda_\beta} {\bf v}^\alpha  ({\bf
  v}^{\alpha T} {\bf D} {\bf v}^\beta) {\bf v}^{\beta T}
$$
into the integral representation for $w_\pm$, one obtains
\begin{eqnarray}
\label{wpm_solution}
&&w_\pm(\bxp,t) =\\
&&= \mp(2\pi)^{-(2d-1)/2}(\det \sigma_\pm)^{-1/2} 
\times \nonumber \\
&& \times \int d\bxp
e^{- {1\over 2}
(\bxp-\bar \bxp(t))\sigma_\pm^{-1}(\bxp-\bar \bxp(t)) } w_{\pm,0}(\bar
\bxp) \nonumber\\
&&=\mp(2\pi)^{-(2d-1)/2} (\det \sigma_\pm)^{-1/2} 
\times \nonumber \\
&& \times \int d\bxp
e^{- {1\over 2}
(\bxp-\bar \bxp)\sigma_\pm^{-1}(\bxp-\bar \bxp) } w_{\pm,0}(\bar
\bxp(-t))\nonumber 
\end{eqnarray}
With this background, we return to the discussion of the behavior
of the generators $w_\pm(\bxp')$ on the reference trajectory. For
times $t<\lambda^{-1}$ (corresponding to energies $\epsilon >
\hbar \lambda$), $\sigma_\pm\approx v_F \lambda_F t\openone$, implying that
only points $\bar \bxp$ in the immediate vicinity of the trajectory
contribute to the integral (\ref{wpm_solution}).
As a result, $w_\pm(\bxp',t)\ll 1$ 
and the Green function remains essentially metallic.  
For $t>\lambda^{-1}$, the component of the matrix $\sigma_+$
corresponding to the largest Liapunov exponent, behaves as
$(\sigma_+)_{00}\sim v_F \lambda_F e^{\lambda t}/\lambda$, and 
the integral begins to sample an exponentially growing region around
the reference trajectory. For 
$$
{v_F \lambda_F \over \lambda} e^{\lambda t}\sim a^2 \Rightarrow t\sim
t_{\rm E}
\Rightarrow \epsilon \sim \hbar/t_{\rm E},
$$
field configurations $w_{+0}(y_i\sim a) = {\cal O}(1)$ contribute to
the integral,
which means that $w_+(\bxp',t>t_{\rm E})= {\cal  O}(1)$. This
identifies the inverse of the Ehrenfest time as the crossover scale
beyond which the Eilenberger Green function, even on exceptionally long 
trajectories, is locked to the superconductor order parameter.

\end{multicols}


\begin{references}
\bibitem{fn1}
The conversion of electrons to holes is accompanied by a phase shift of
  $i\exp(i\phi)$, and the reciprocal conversion of holes into electrons by
  $i\exp(-i\phi)$. A trajectory containing a twofold conversion, electron $\to$
  hole $\to$ electron, therefore comes with a total phase shift of $i^2 = -1 =
  \exp(i\hbar^{-1} h/2)$.
\bibitem{fn4}
The quasiclassical Green function is related to the impurity averaged Gorkov
  Green function $G$ through $g(\br,\bn) \approx G(\br,\bn)$, where $$
  G(\br,\bn) \equiv \frac{i}{\pi}\int d\xi\int d\br' e^{ip\bn\cdot
  \br'/\hbar}\left\langle G\left(\br+\frac{\br'}{2}, \br-\frac{\br'}{2} \right)
  \right\rangle, $$ and $G(\br,\br')$ is the microscopic Gorkov Green function.
  Here $\xi =v_F(p-p_F)$. The origin of the only approximate equality between
  the quasiclassical $g(\br,\bn)$ and the $p$-integrated exact Wigner transform
  of the Gorkov Green function $G(\br,\bn)$ is discussed in Appendix
  \ref{app:ball}.
\bibitem{fn5}
We note that this modelling scheme is almost certainly incomplete as it
  suggests that a given realization of a chaotic system auto-regularizes
  itself. In contrast, other approaches to the problem (cf. e.g.
  Ref.\protect\cite{zirnbauer99}) indicate that some minimal averaging over a
  stochastic set of parameters is needed to stabilize the very formulation of a
  quasiclassical Green function approach.
\end{references}
\end{document}